\UseRawInputEncoding
\documentclass[a4paper,aps,prd,10pt,preprintnumbers,showpacs,twocolumn,superscriptaddress,nofootinbib,amsmath,amssymb,floatfix]{revtex4-1}
\usepackage{graphicx}
\usepackage{cmap}
\usepackage[utf8]{inputenc}
\usepackage[T1]{fontenc}

\def\imo{i}
\def\re#1{Re(#1)}
\def\im#1{Im(#1)}
\def\K{{\cal K}}
\def\Order#1{{\cal O}\left(#1\right)}
\newcommand{\added}[1]{#1}

\begin{document}

\title{Scattering of scalar, electromagnetic, and Dirac fields in an asymptotically flat regular black hole supported by primordial dark matter}

\author{S. V. Bolokhov}
\email{bolokhov-sv@rudn.ru}
\affiliation{Peoples’ Friendship University of Russia (RUDN University), 6 Miklukho-Maklaya Street, Moscow 117198, Russia}

\begin{abstract}
We study grey-body factors and absorption cross sections for massless scalar, electromagnetic, and Dirac fields in the exact asymptotically flat regular black-hole geometry supported by a phantom Dirac--Born--Infeld scalar. In the black-hole branch all three sectors are governed by single-barrier effective potentials, which allows a direct 6th-order WKB treatment of the scattering problem and a comparison with the recent quasinormal-mode/grey-body-factor correspondence. We show that increasing the regularity parameter lowers the barriers, shifts transmission to lower frequencies, and enhances the absorption cross sections in all three sectors. By comparing the direct WKB grey-body factors with those reconstructed from the lowest quasinormal modes, we explicitly test the QNM/GBF correspondence and find good agreement, typically at the level of $10^{-2}$ or better.
\end{abstract}

\maketitle

\section{Introduction}

Grey-body factors quantify how the effective curvature potential surrounding a compact object filters radiation propagating between the horizon and spatial infinity. For black holes they determine the deviation from a purely black-body Hawking spectrum \cite{Hawking:1975vcx} and, more generally, characterize the transmission and reflection properties of waves scattered by the spacetime geometry \cite{Page:1976df,Page:1976ki,Kanti:2002nr}. In some situations they can influence the Hawking emission even more strongly than the temperature itself, because the lowering of the effective barrier may overcompensate the cooling of the black hole \cite{Konoplya:2019ppy}. Because they depend directly on the detailed shape of the effective potential, grey-body factors provide information complementary to quasinormal modes and are an efficient probe of near-horizon physics, asymptotic structure, and spin-dependent propagation effects \cite{Kokkotas:1999bd,Berti:2009kk,Konoplya:2011qq,Konoplya:2024gbf,Konoplya:2024vuj}.

Regular black holes are especially interesting from this perspective. Replacing the central singularity by a smooth core modifies the effective potential while preserving the black-hole character of the exterior region, so the corresponding transmission rates can differ in a systematic way from their Schwarzschild counterparts. A broad literature has examined quasinormal modes, Hawking radiation, and grey-body factors for regular geometries in a variety of models \cite{Flachi:2012nv,Yang:2021cvh,Konoplya:2022hll,Cai:2021ele,Li:2014fka,Guo:2024jhg,Konoplya:2023aph,Jusufi:2020odz,Huang:2023aet,Lopez:2022uie,Konoplya:2023ahd,Fernando:2012yw,MahdavianYekta:2019pol,Gingrich:2024tuf,Konoplya:2026gim,Konoplya:2025ect,Jawad:2020hju,Pedraza:2021hzw,Bonanno:2025dry,Lin:2013ofa,Konoplya:2023ppx,Held:2019xde,Arbelaez:2025gwj,Arbelaez:2026eaz,Konoplya:2025uta,Skvortsova:2024wly,Skvortsova:2024eqi,Bolokhov:2025fto,Malik:2026lfj,Saka:2025xxl,Malik:2025qnr,Malik:2024tuf,Malik:2024itg,Dubinsky:2026gcj,Dubinsky:2026wcv,Dubinsky:2025wns}. In the present case this issue is particularly appealing because the underlying geometry is not introduced phenomenologically, but follows from an exact matter-supported solution.
It has been suggested that, even within broad parametrized frameworks for black hole geometries, strong-field observables such as quasinormal frequencies or grey-body factors are largely controlled by a limited set of effective parameters  \cite{Konoplya:2020pbh}. In this light, isolating and analyzing the influence of the single regularization scale $a$ in the present exact solution provides a natural and well-justified way to capture the dominant physical effects.

Recently, Parvez and Shankaranarayanan obtained an exact asymptotically flat, nonsingular black-hole solution sourced by a phantom Dirac--Born--Infeld (DBI) scalar \cite{Parvez:2025dbi}. The solution is regular at the center, carries scalar hair, and interpolates between black-hole, extremal-remnant, and horizonless regular configurations depending on the parameters \cite{Parvez:2025dbi}. Perturbations and spectra of a massive scalar field in this background has been recently considered in \cite{Skvortsova:2026jtx}, while the quasinormal modes of  massless test fields have been studied in \cite{Lutfuoglu:2026boa}. In this paper we analyze grey-body factors and absorption cross sections for scalar, electromagnetic, and Dirac fields in the DBI-supported regular black-hole spacetime. After reviewing the geometry and the effective potentials, we compute the transmission probabilities by the 6th-order WKB method, compare them with the reconstruction based on the lowest quasinormal frequencies, use the resulting grey-body factors to evaluate the corresponding absorption cross sections, and then add a compact geometric-optics estimate of the associated Hawking emission. 

The paper is organized as follows. In Sec.~II we review the DBI-supported regular black-hole geometry. In Sec.~III we derive the effective potentials for the scalar, electromagnetic, and Dirac sectors and discuss their qualitative dependence on the regularity parameter and the multipole number. In Sec.~IV we analyze the grey-body factors obtained with the 6th-order WKB method, assess the accuracy of the QNM/GBF correspondence, discuss the corresponding absorption cross sections, and present a compact geometric-optics estimate of the Hawking emission. Finally, in Sec.~V we summarize the main results and outline possible extensions of the present work.

\section{Background geometry}\label{sec:geometry}

Following Ref.~\cite{Parvez:2025dbi}, we consider Einstein gravity minimally coupled to a non-linear Dirac--Born--Infeld scalar field. A convenient form of the action is
\begin{equation}
S=\int d^4x\sqrt{-g}\left[\frac{R}{16\pi G}-\epsilon\Lambda^4\left(\sqrt{1+\frac{\nabla_\mu\phi\nabla^\mu\phi}{\Lambda^4}}-1\right)\right],
\end{equation}
where $\Lambda$ sets the DBI scale and $\epsilon=+1$ or $-1$ distinguishes the canonical and phantom branches. The regular black-hole solution relevant here is supported by the phantom branch, $\epsilon=-1$ \cite{Parvez:2025dbi}.

The exact static, spherically symmetric line element can be written as
\begin{equation}\label{eq:metric}
ds^2=-f(r)dt^2+\frac{dr^2}{f(r)}+R^2(r)(d\theta^2+\sin^2\theta\,d\phi^2),
\end{equation}
with
\begin{equation}\label{eq:metricfunctions}
\begin{aligned}
f(r)&=1+\frac{3M}{a}\left(\frac{r}{a}-\frac{a^2+r^2}{a^2}\arctan\frac{a}{r}\right),\\
R^2(r)&=r^2+a^2.
\end{aligned}
\end{equation}
Here $M$ is the ADM mass and $a$ is the regularity scale. The areal radius never shrinks below $a$, so the central singularity is replaced by a regular minimal two-sphere. At large distances one has $R(r)\sim r$ and
\begin{equation}
f(r)=1-\frac{2M}{r}+\Order{r^{-3}},
\end{equation}
which shows that the solution is asymptotically flat. Throughout the paper we set $M=1$.

The tortoise coordinate is defined by
\begin{equation}\label{eq:tortoise}
dr_*\equiv\frac{dr}{f(r)}.
\end{equation}
For the black-hole branch, $r_*\to -\infty$ at the event horizon and $r_*\to +\infty$ at spatial infinity. This coordinate is the natural one for casting the perturbation equations into a Schr\"odinger-type form and for formulating the associated scattering problem.

\section{Test fields and effective potentials}\label{sec:fields}

For each spin sector considered here, the field equations reduce to a one-dimensional wave equation,
\begin{equation}\label{eq:mastergeneral}
\frac{d^2\Psi}{dr_*^2}+\left(\omega^2-V(r)\right)\Psi=0,
\end{equation}
where the effective potential $V(r)$ depends on the spin of the field.

\subsection{Scalar field}

The massless scalar field obeys the Klein--Gordon equation
\begin{equation}
\Box\Phi=0.
\end{equation}
Using the decomposition
\begin{equation}
\Phi(t,r,\theta,\phi)=e^{-\imo\omega t}Y_{\ell m}(\theta,\phi)\frac{\Psi_s(r)}{R(r)},
\end{equation}
one finds Eq.~(\ref{eq:mastergeneral}) with \cite{Carter1968HJ,Carter1968Kerr,Konoplya:2018arm}
\begin{equation}\label{eq:Vscalar}
V_s(r)=f(r)\left[\frac{\ell(\ell+1)}{R^2(r)}+\frac{1}{R(r)}\frac{d^2R(r)}{dr_*^2}\right],
\end{equation}
where $\ell=0,1,2,\ldots$. For the DBI-supported geometry this can be written more explicitly as
\begin{equation}
V_s(r)=f\left[\frac{\ell(\ell+1)}{r^2+a^2}+\frac{r f'}{r^2+a^2}+\frac{a^2f}{(r^2+a^2)^2}\right].
\end{equation}

\subsection{Electromagnetic field}

After decomposition into vector spherical harmonics, both parity sectors of the Maxwell field reduce to Eq.~(\ref{eq:mastergeneral}) with the effective potential \cite{Ruffini:1973}
\begin{equation}\label{eq:Vem}
V_{em}(r)=f(r)\frac{\ell(\ell+1)}{R^2(r)}=f(r)\frac{\ell(\ell+1)}{r^2+a^2},
\end{equation}
valid for $\ell=1,2,3,\ldots$. This potential is purely centrifugal and does not contain the additional curvature term present in the scalar case.

\subsection{Dirac field}

For the massless Dirac field, the radial equations can be reduced to two supersymmetric partner equations of the form (\ref{eq:mastergeneral}) for the combinations $Z_{\pm}=F\pm G$, with
\begin{equation}\label{eq:Vdiraccompact}
V_{\pm}(r)=W^2\pm\frac{dW}{dr_*},
\end{equation}
where the superpotential is
\begin{equation}
W(r)=\frac{|\kappa|\sqrt{f(r)}}{R(r)}=\frac{|\kappa|\sqrt{f(r)}}{\sqrt{r^2+a^2}},
\end{equation}
and $\kappa=\pm 1,\pm 2,\ldots$ is the usual separation constant. An explicit form is \cite{Brill:1957fx}
\begin{equation}\label{eq:Vdirac}
V_{\pm}(r)=\frac{\kappa^2f}{r^2+a^2}\pm |\kappa|\sqrt{f}\left[\frac{f'}{2\sqrt{r^2+a^2}}-\frac{fr}{(r^2+a^2)^{3/2}}\right].
\end{equation}
The two partner potentials are isospectral, so one may work with either of them when computing the transmission coefficients.

For the black-hole branch all three sectors are described by single-barrier effective potentials which vanish at both asymptotic ends. This is precisely the setting in which standard one-barrier WKB scattering methods are expected to work reliably.

\begin{figure*}[t]
\centering
\includegraphics[width=0.98\textwidth]{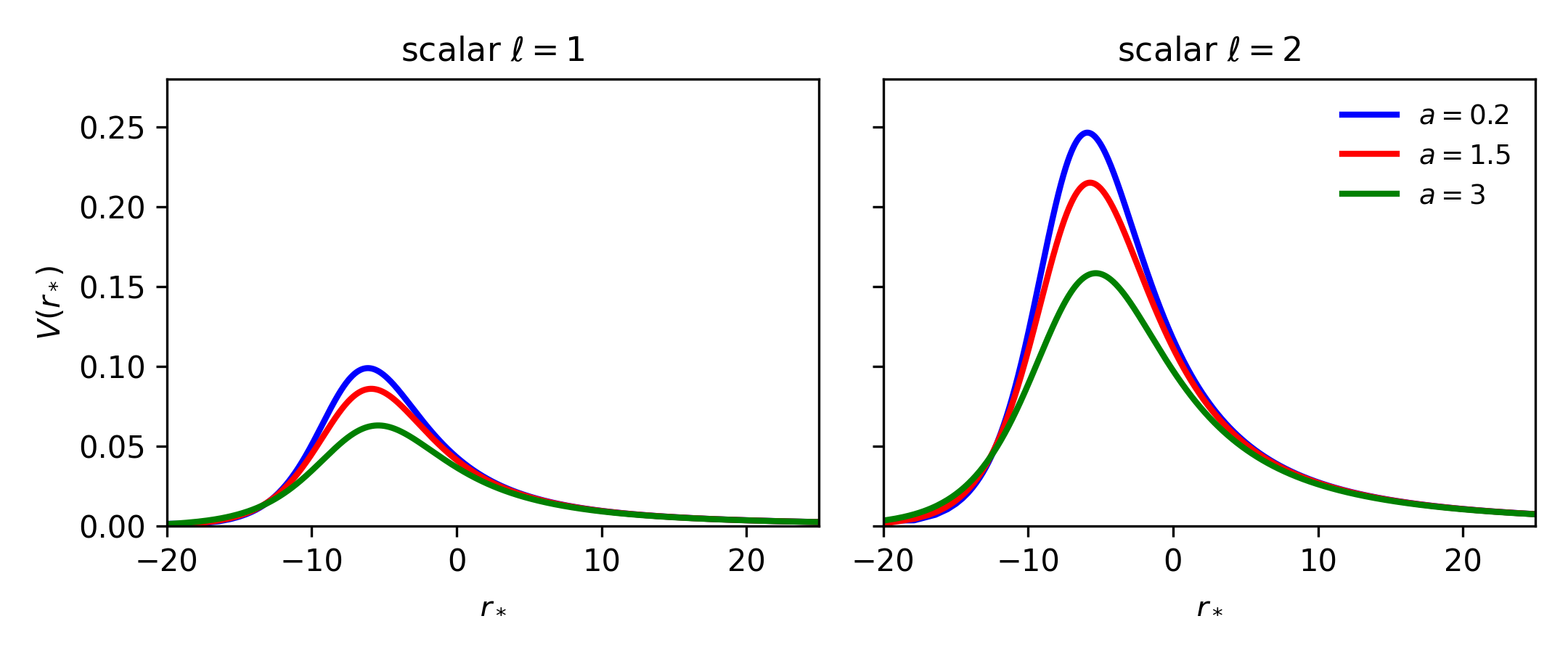}
\caption{\added{Effective scalar potentials $V_s(r_*)$ for the same parameter values used in the grey-body-factor analysis, with $M=1$. Left panel: $\ell=1$. Right panel: $\ell=2$. In each panel the curves correspond to $a=0.2$ (blue), $a=1.5$ (red), and $a=3$ (green). Increasing $a$ lowers and broadens the barrier, whereas increasing $\ell$ raises its peak.}}
\label{fig:potential_scalar_gbf}
\end{figure*}

\begin{figure*}[t]
\centering
\includegraphics[width=0.55\textwidth]{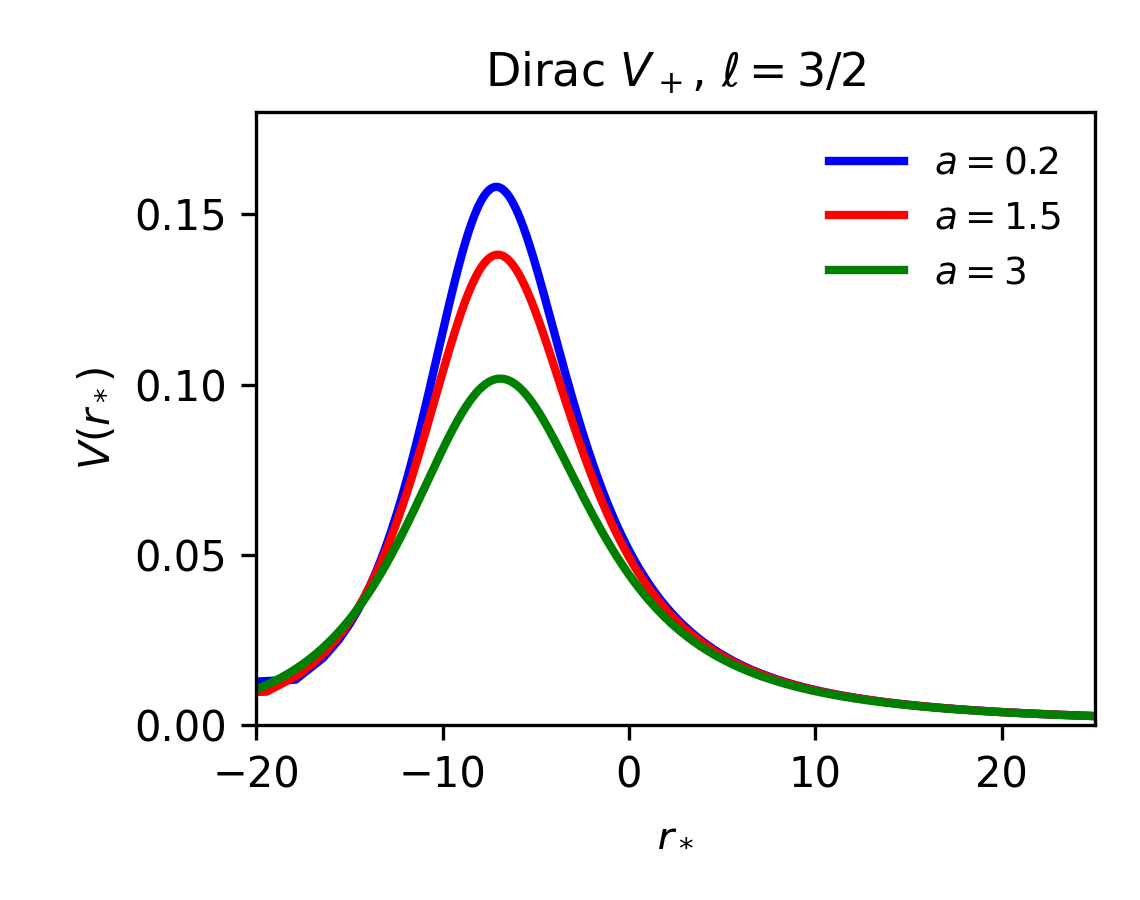}
\caption{\added{Effective Dirac partner potential $V_+(r_*)$ for $\ell=3/2$ (equivalently $|\kappa|=2$) and $M=1$. The curves correspond to $a=0.2$ (blue), $a=1.5$ (red), and $a=3$ (green). The barrier height decreases systematically as the regularity parameter grows.}}
\label{fig:potential_dirac_gbf}
\end{figure*}

\begin{figure*}[t]
\centering
\includegraphics[width=0.98\textwidth]{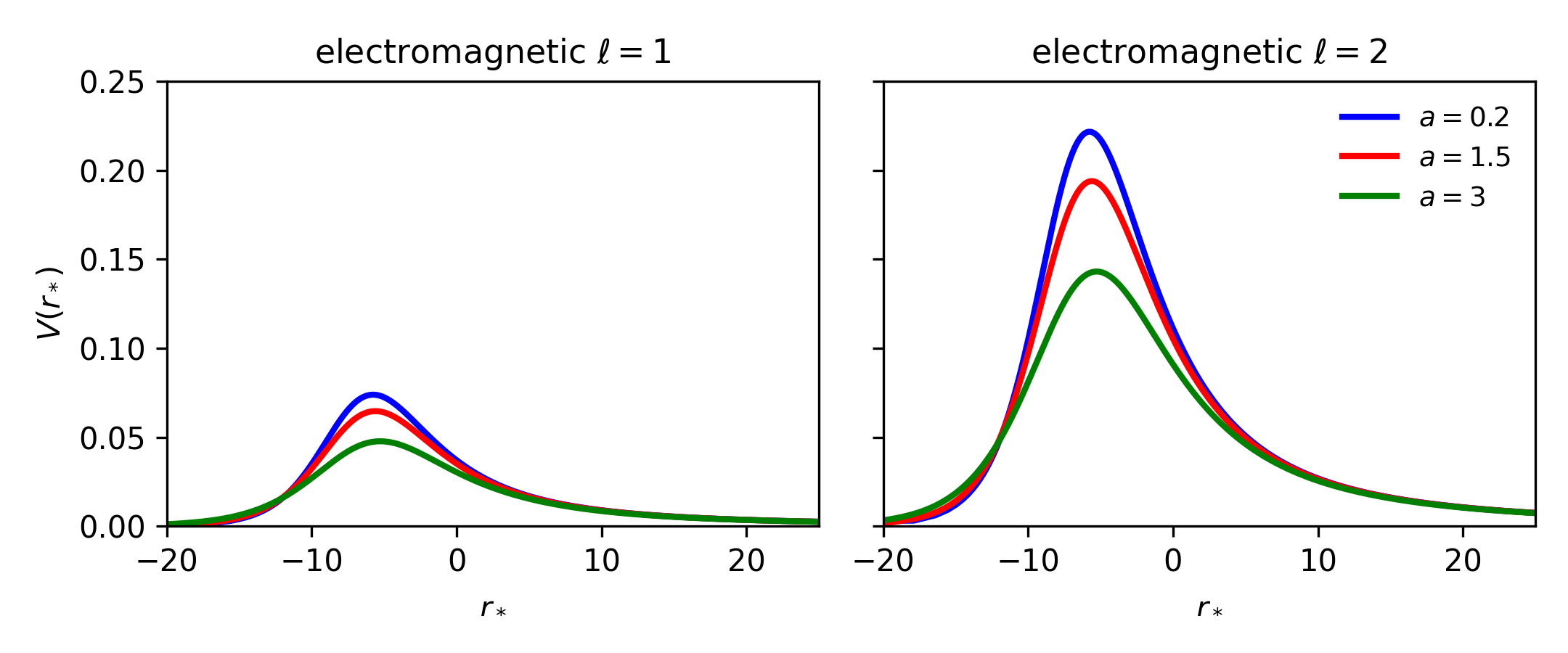}
\caption{\added{Effective electromagnetic potentials $V_{em}(r_*)$ for the same parameter values used in the grey-body-factor analysis, with $M=1$. Left panel: $\ell=1$. Right panel: $\ell=2$. In each panel the curves correspond to $a=0.2$ (blue), $a=1.5$ (red), and $a=3$ (green). As in the scalar case, increasing $a$ lowers the barrier, while increasing $\ell$ raises it through the centrifugal term.}}
\label{fig:potential_electromagnetic_gbf}
\end{figure*}

\added{For the same values of the regularity parameter and multipole numbers that will be used below in the grey-body-factor analysis, the corresponding effective barriers are shown in Figs.~\ref{fig:potential_scalar_gbf}--\ref{fig:potential_electromagnetic_gbf}. At fixed multipole, increasing $a$ lowers the maximum of the barrier and makes the profile broader in all three sectors. Since a lower barrier reflects less efficiently, this deformation is expected to enhance the grey-body factors and to shift their onset to smaller frequencies.}

\added{The dependence on the angular number is especially transparent in the bosonic sectors: increasing $\ell$ from $1$ to $2$ raises the centrifugal contribution in both the scalar and electromagnetic potentials, so the $\ell=2$ barriers are higher than the $\ell=1$ ones and the corresponding transmission probabilities should be smaller at the same frequency. For the Dirac field we display the partner potential $V_+$ for $\ell=3/2$, which corresponds to $|\kappa|=2$; here again the systematic decrease of the barrier with growing $a$ anticipates the increase of the transmission rate.}

\section{Grey-body factors and absorption cross sections}\label{sec:gbf}

\subsection{Grey-body factors and the QNM--GBF correspondence}

For a wave incident from spatial infinity, the asymptotic solutions of Eq.~(\ref{eq:mastergeneral}) can be written as
\begin{equation}\label{eq:scattbc}
\Psi\sim
\begin{cases}
T(\omega)e^{-\imo\omega r_*}, & r_*\to -\infty,\\
 e^{-\imo\omega r_*}+R(\omega)e^{+\imo\omega r_*}, & r_*\to +\infty,
\end{cases}
\end{equation}
where $T(\omega)$ and $R(\omega)$ are the transmission and reflection amplitudes, respectively. For real frequencies, the conserved Wronskian implies
\begin{equation}
|R(\omega)|^2+|T(\omega)|^2=1.
\end{equation}
The partial grey-body factor is then defined as the transmission probability,
\begin{equation}\label{eq:Gamma}
\Gamma_{\ell}(\omega)=|T_{\ell}(\omega)|^2,
\end{equation}
with the appropriate replacement of $\ell$ by $|\kappa|$ in the Dirac case.

When the effective potential has a single smooth maximum, the transmission coefficient can be computed semi-analytically by the WKB approach. In this method one matches asymptotic WKB solutions across the turning points and expresses the scattering coefficients in terms of the local behavior of the potential near its peak \cite{Iyer:1986np,Konoplya:2003ii,Matyjasek:2017psv,Konoplya:2019hlu}.

In the standard notation, the WKB result for the transmission probability can be written as
\begin{equation}\label{eq:wkbGamma}
\Gamma_{\ell}(\omega)=|T_{\ell}(\omega)|^2=\left(1+e^{2\pi i\K_{\ell}(\omega)}\right)^{-1},
\end{equation}
where $\K_{\ell}(\omega)$ is the WKB quantity determined by the effective potential and its derivatives at the maximum. At leading orders one has the schematic structure
\begin{equation}
\K=\frac{i\bigl(\omega^2-V_0\bigr)}{\sqrt{-2V_2}}+\sum_{j=2}^{N}\Lambda_j,
\end{equation}
where $V_0$ is the value of the potential at the peak, $V_2$ is the second derivative with respect to $r_*$ at that point, and the higher-order corrections $\Lambda_j$ depend on successive derivatives of the potential. WKB methods at various orders have been used extensively in studies of black holes and other compact objects, where they generally show very good agreement with more accurate approaches whenever the problem is governed by a single smooth barrier and the multipole number is not smaller than the overtone number of the quasinormal mode \cite{Konoplya:2010vz,Malik:2025erb,Guo:2020caw,Kokkotas:2010zd,Konoplya:2009hv,Stuchlik:2025mjj,Skvortsova:2024msa,Abdalla:2005hu,Bolokhov:2025lnt,Konoplya:2006ar,Pathrikar:2025gzu,Ishihara:2008re,Bolokhov:2025aqy,Fernando:2016ftj,Momennia:2018hsm,Eniceicu:2019npi,Wongjun:2019ydo,Breton:2017hwe,Karmakar:2023cwg,Konoplya:2001ji,Konoplya:2005sy,Bolokhov:2025egl,Lutfuoglu:2025ljm,Lutfuoglu:2026gey,Lutfuoglu:2025ohb,Lutfuoglu:2026gis,Malik:2024iky,Malik:2024sxv}.
At the same time, for potential having double-well shape the WKB method is highly inaccurate and numerical integration must be used instead \cite{Konoplya:2025ixm}.

For the DBI-supported regular black hole this approach is appropriate because the effective potentials in all three massless sectors are single barriers in the black-hole regime. The numerical implementation therefore consists of the following steps: identify the maximum of the effective potential for fixed spin and multipole number, compute the required derivatives with respect to $r_*$, evaluate the WKB quantity $\K_{\ell}(\omega)$, and then obtain the grey-body factor from Eq.~(\ref{eq:wkbGamma}). This procedure can be repeated for different values of the regularity parameter $a$, frequency $\omega$, and multipole number in order to map the spin dependence of the transmission probabilities.

A particularly useful aspect of the present setup is its relation to the quasinormal spectrum. Recent work has shown that, for single-barrier potentials, one can reconstruct the WKB scattering quantity entering the grey-body factor from the lowest quasinormal frequencies, beginning with the eikonal approximation and improving it by including the first overtone \cite{Konoplya:2024gbf,Konoplya:2024vuj}. This correspondence provides an additional consistency check and a useful bridge between the ringdown and scattering descriptions.

More specifically, the correspondence is useful because it translates spectral information into scattering data: once the effective potential is a single smooth barrier, the same local geometry of the peak that controls the WKB quasinormal spectrum also determines the transmission probability. Although the present DBI background is spherically symmetric, the correspondence itself is more general: it was first derived for asymptotically flat spherically symmetric black holes and then extended to axially symmetric black holes admitting separation of variables, at least in the nonsuperradiant regime \cite{Konoplya:2024gbf,Konoplya:2024vuj}. Writing $\omega_0\equiv\omega_{\ell 0}$ and $\omega_1\equiv\omega_{\ell 1}$, the leading eikonal relation is
\begin{equation}\label{eq:gbf_qnm_eikonal}
i\K_{\ell}(\omega)=\frac{\omega^2-\re{\omega_0}^2}{4\,\re{\omega_0}\,\im{\omega_0}}+\Order{\ell^{-1}},
\end{equation}
while the improved formula corresponding to Eq.~(23) of Ref.~\cite{Konoplya:2024vuj} can be written as
\begin{widetext}
\begin{equation}
%\resizebox{\columnwidth}{!}{$\displaystyle
\begin{aligned}
i\K_{\ell}(\omega)&=\frac{\omega^2-\re{\omega_0}^2}{4\,\re{\omega_0}\,\im{\omega_0}}
\left(1+\frac{\left(\re{\omega_0}-\re{\omega_1}\right)^2}{32\,\im{\omega_0}^2}
-\frac{3\,\im{\omega_0}-\im{\omega_1}}{24\,\im{\omega_0}}\right)\\
&\quad -\frac{\left(\omega^2-\re{\omega_0}^2\right)^2}{16\,\re{\omega_0}^3\,\im{\omega_0}}
\left(1+\frac{\re{\omega_0}\left(\re{\omega_0}-\re{\omega_1}\right)}{4\,\im{\omega_0}^2}\right)\\
&\quad +\frac{\left(\omega^2-\re{\omega_0}^2\right)^3}{32\,\re{\omega_0}^5\,\im{\omega_0}}\times
\left(1+\frac{\re{\omega_0}\left(\re{\omega_0}-\re{\omega_1}\right)}{4\,\im{\omega_0}^2}
+\frac{\re{\omega_0}^2\left(\re{\omega_0}-\re{\omega_1}\right)^2}{16\,\im{\omega_0}^4}\right.\\
&\qquad \left.-\frac{3\,\im{\omega_0}-\im{\omega_1}}{12\,\im{\omega_0}}\right)+\Order{\ell^{-3}}.
\end{aligned}
%$}
\end{equation}
\end{widetext}
This improved relation provides a more accurate estimate of the WKB scattering quantity entering Eq.~(\ref{eq:wkbGamma}) and has already been tested in a number of black-hole and regular-geometry settings, generally showing good agreement even for moderately low multipole numbers \cite{Konoplya:2010vz,Malik:2024wvs,
Dubinsky:2024vbn,Lutfuoglu:2025eik,Lutfuoglu:2025kqp,
Lutfuoglu:2025mqa,Lutfuoglu:2025ldc,
Lutfuoglu:2026uzy, Bolokhov:2026eqf, Bolokhov:2025lnt, Bolokhov:2024otn, Skvortsova:2024msa,Malik:2025dxn,Malik:2024cgb}. At the same time, the correspondence inherits the limitations of the WKB picture itself: if the effective potential ceases to be a single barrier, for example by developing a double-well structure \cite{Konoplya:2025hgp}, or if higher-curvature effects strongly deform the centrifugal sector and induce catastrophic instabilities \cite{Konoplya:2017zwo,Takahashi:2010gz,Dotti:2004sh,Konoplya:2017lhs,Konoplya:2017ymp}, the reconstruction of grey-body factors from quasinormal frequencies is expected to lose accuracy or fail.

\begin{figure*}[t]
\centering
\includegraphics[width=0.48\textwidth]{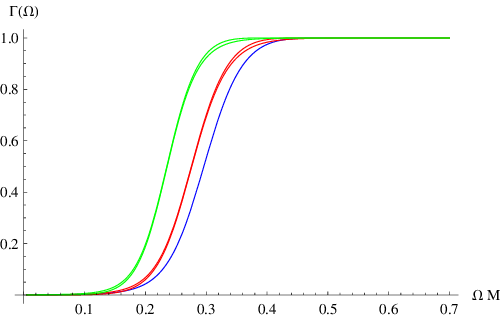}\hfill
\includegraphics[width=0.48\textwidth]{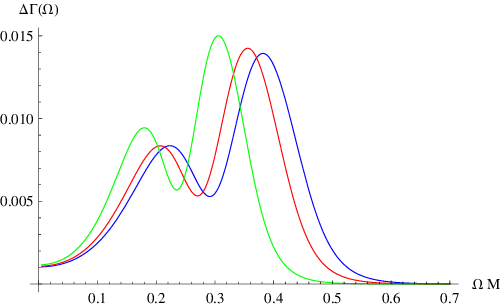}
\caption{Comparison between the grey-body factors computed by the 6th-order WKB method and by the correspondence based on Eq.~(\ref{eq:gbf_qnm_eikonal}) for the scalar field with $\ell=1$ and $M=1$, with $a=0.2$ (blue), $a=1.5$ (red), and $a=3$ (green). Left panel: grey-body factors $\Gamma(\omega)$ obtained with the two approaches. Right panel: difference $\Delta\Gamma(\omega)$ between the corresponding grey-body factors for the same values of the regularity parameter.}
\label{fig:gbf_correspondence_comparison}
\end{figure*}

\begin{figure*}[t]
\centering
\includegraphics[width=0.48\textwidth]{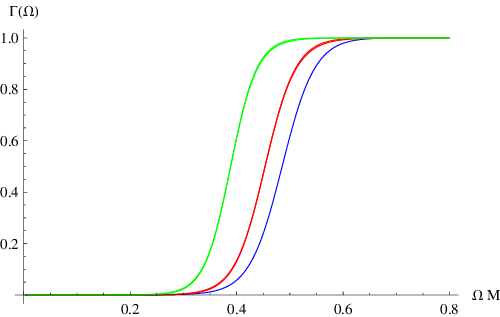}\hfill
\includegraphics[width=0.48\textwidth]{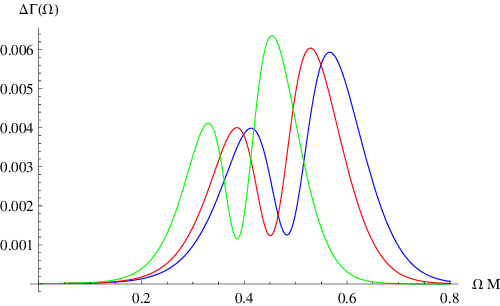}
\caption{Comparison between the grey-body factors computed by the 6th-order WKB method and by the correspondence based on Eq.~(\ref{eq:gbf_qnm_eikonal}) for the scalar field with $\ell=2$ and $M=1$, with $a=0.2$ (blue), $a=1.5$ (red), and $a=3$ (green). Left panel: grey-body factors $\Gamma(\omega)$ obtained with the two approaches. Right panel: difference $\Delta\Gamma(\omega)$ between the corresponding grey-body factors for the same values of the regularity parameter.}
\label{fig:gbf_correspondence_comparison_l2}
\end{figure*}

\begin{figure*}[t]
\centering
\includegraphics[width=0.48\textwidth]{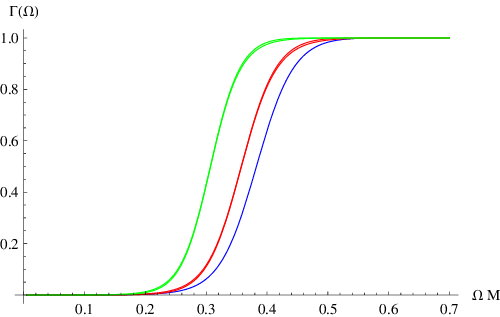}\hfill
\includegraphics[width=0.48\textwidth]{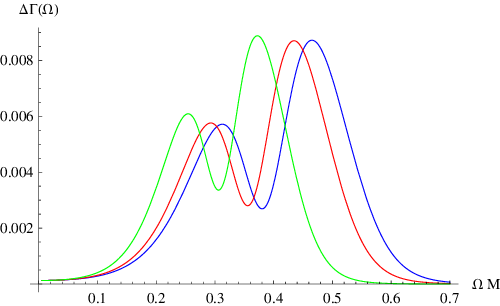}
\caption{Comparison between the grey-body factors computed by the 6th-order WKB method and by the correspondence based on Eq.~(\ref{eq:gbf_qnm_eikonal}) for the Dirac field with $\ell=3/2$ and $M=1$, with $a=0.2$ (blue), $a=1.5$ (red), and $a=3$ (green). Left panel: grey-body factors $\Gamma(\omega)$ obtained with the two approaches. Right panel: difference $\Delta\Gamma(\omega)$ between the corresponding grey-body factors for the same values of the regularity parameter.}
\label{fig:gbf_correspondence_comparison_dirac_l3half}
\end{figure*}

\begin{figure*}[t]
\centering
\includegraphics[width=0.48\textwidth]{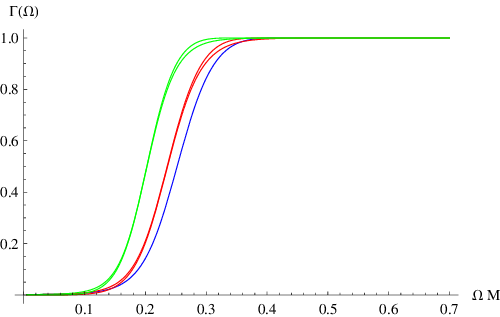}\hfill
\includegraphics[width=0.48\textwidth]{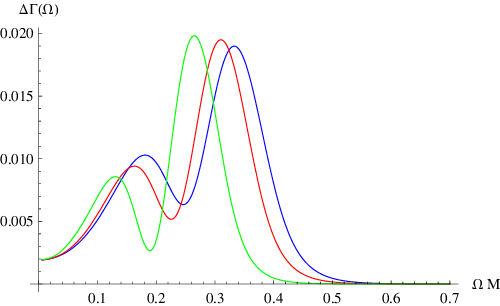}
\caption{Comparison between the grey-body factors computed by the 6th-order WKB method and by the correspondence based on Eq.~(\ref{eq:gbf_qnm_eikonal}) for the electromagnetic field with $\ell=1$ and $M=1$, with $a=0.2$ (blue), $a=1.5$ (red), and $a=3$ (green). Left panel: grey-body factors $\Gamma(\omega)$ obtained with the two approaches. Right panel: difference $\Delta\Gamma(\omega)$ between the corresponding grey-body factors for the same values of the regularity parameter.}
\label{fig:gbf_correspondence_comparison_em_l1}
\end{figure*}

\begin{figure*}[t]
\centering
\includegraphics[width=0.48\textwidth]{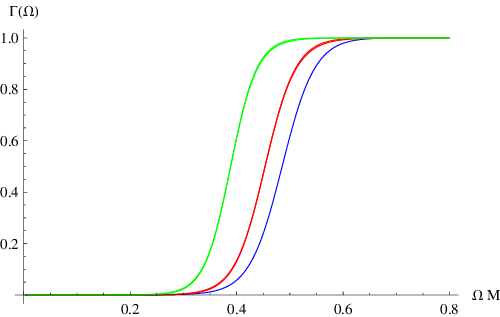}\hfill
\includegraphics[width=0.48\textwidth]{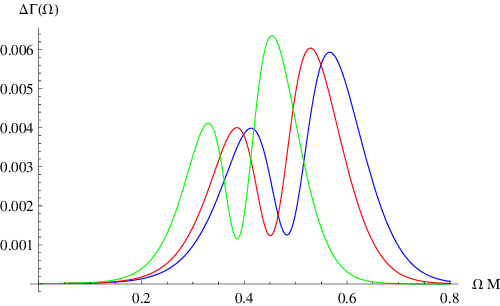}
\caption{Comparison between the grey-body factors computed by the 6th-order WKB method and by the correspondence based on Eq.~(\ref{eq:gbf_qnm_eikonal}) for the electromagnetic field with $\ell=2$ and $M=1$, with $a=0.2$ (blue), $a=1.5$ (red), and $a=3$ (green). Left panel: grey-body factors $\Gamma(\omega)$ obtained with the two approaches. Right panel: difference $\Delta\Gamma(\omega)$ between the corresponding grey-body factors for the same values of the regularity parameter.}
\label{fig:gbf_correspondence_comparison_em_l2}
\end{figure*}

\added{To assess how accurately this correspondence is fulfilled in the present geometry, Figs.~\ref{fig:gbf_correspondence_comparison}--\ref{fig:gbf_correspondence_comparison_em_l2} compare the grey-body factors obtained by the direct 6th-order WKB calculation with those reconstructed from the leading QNM/GBF relation, Eq.~(\ref{eq:gbf_qnm_eikonal}).}

\added{Several features are immediate. In all left panels the curves obtained from the direct WKB calculation and from the quasinormal-mode correspondence are nearly indistinguishable, so the correspondence reproduces not only the qualitative ordering of the transmission curves but also their quantitative location along the frequency axis. For fixed spin and multipole number, increasing the regularity parameter from $a=0.2$ to $a=3$ shifts the transition region to smaller $\omega$, in agreement with the lowering of the effective barriers displayed in Figs.~\ref{fig:potential_scalar_gbf}--\ref{fig:potential_electromagnetic_gbf}. Thus the DBI regularity scale makes the black hole more transparent to incoming radiation in all three massless sectors.}

\added{The right panels show that the absolute difference $\Delta\Gamma$ is concentrated in the crossover region where $0<\Gamma<1$, while it becomes negligible in the low- and high-frequency limits where both approaches give $\Gamma\simeq0$ or $\Gamma\simeq1$. This localization is expected for a single-barrier problem: once the transmission curve has saturated, a small mismatch in the reconstructed WKB quantity produces only a very small change in $\Gamma$. Even at the leading eikonal level, the maximal discrepancy remains of order $10^{-2}$ or smaller for the representative modes shown here; the largest mismatch occurs for the electromagnetic mode with $\ell=1$, while the scalar and electromagnetic modes with $\ell=2$ already reduce the peak error to a few $10^{-3}$. The Dirac mode with $\ell=3/2$ lies between these two limits. Therefore the correspondence is fulfilled with good quantitative accuracy already for moderate multipole numbers, and the expected improvement toward the eikonal regime is clearly visible. Including the first overtone through the improved relation above should sharpen the agreement even further, especially for the lowest multipoles.}

\subsection{Absorption cross sections}

Once the transmission probabilities are known, one may construct observable quantities such as the absorption cross section and the frequency-resolved Hawking flux. For the massless scalar field, the partial absorption cross section associated with a fixed multipole number is
\begin{equation}
\sigma^{(s)}_{\ell}(\omega)=\frac{\pi}{\omega^2}(2\ell+1)\Gamma_{\ell}(\omega),
\end{equation}
where the factor $(2\ell+1)$ is the usual spherical-harmonic degeneracy. Summing over all multipoles, one obtains the total scalar absorption cross section,
\begin{equation}
\sigma^{(s)}_{\rm abs}(\omega)=\sum_{\ell=0}^{\infty}\sigma^{(s)}_{\ell}(\omega)=\frac{\pi}{\omega^2}\sum_{\ell=0}^{\infty}(2\ell+1)\Gamma_{\ell}(\omega).
\end{equation}
Analogous expressions apply in the electromagnetic and Dirac sectors, with the corresponding spin-dependent multiplicities \cite{Page:1976df,Page:1976ki,Kanti:2002nr,Konoplya:2011qq,Konoplya:2024gbf,Konoplya:2024vuj}. Because each partial absorption cross section contains the universal factor $\pi/\omega^2$, it does not remain a monotonic sigmoid even though the corresponding grey-body factor does. Instead, each partial wave develops a peak near the frequency range where its transmission probability turns on most rapidly, and then decreases again at larger $\omega$. The figures below are obtained by summing the first 50 partial waves, which is sufficient to capture the displayed total cross sections.

\begin{figure*}[t]
\centering
\includegraphics[width=0.48\textwidth]{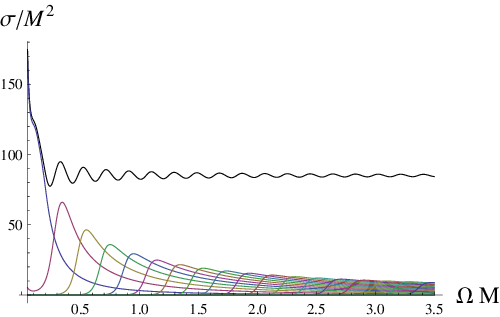}\hfill
\includegraphics[width=0.48\textwidth]{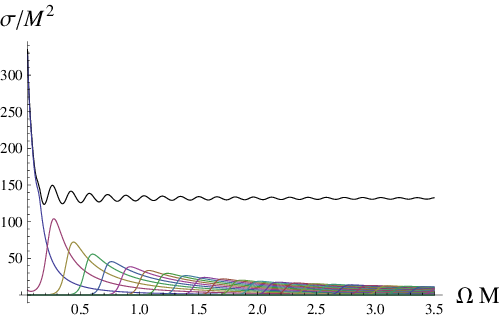}
\caption{Partial and total scalar absorption cross sections for the DBI-supported regular black hole at $M=1$, obtained by summing the first 50 multipoles. Left panel: absorption cross sections for $a=0.2$. Right panel: absorption cross sections for $a=3$. In each panel, the partial absorption cross sections are shown together with the corresponding total absorption cross section.}
\label{fig:scalar_absorption_compare}
\end{figure*}

\begin{figure*}[t]
\centering
\includegraphics[width=0.48\textwidth]{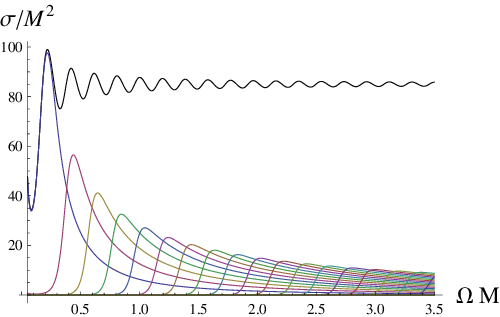}\hfill
\includegraphics[width=0.48\textwidth]{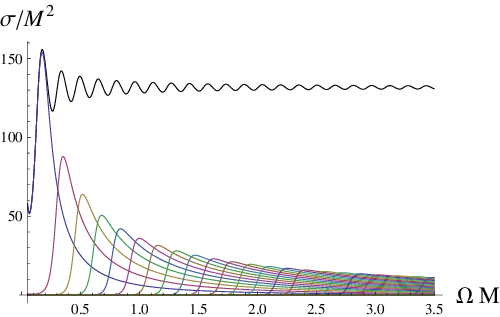}
\caption{Partial and total Dirac absorption cross sections for the DBI-supported regular black hole at $M=1$, obtained by summing the first 50 multipoles. Left panel: absorption cross sections for $a=0.2$. Right panel: absorption cross sections for $a=3$. In each panel, the partial absorption cross sections are shown together with the corresponding total absorption cross section.}
\label{fig:dirac_absorption_compare}
\end{figure*}

\begin{figure*}[t]
\centering
\includegraphics[width=0.48\textwidth]{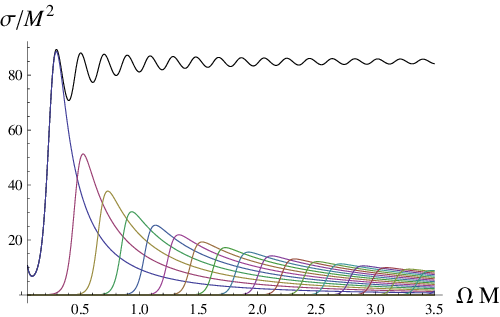}\hfill
\includegraphics[width=0.48\textwidth]{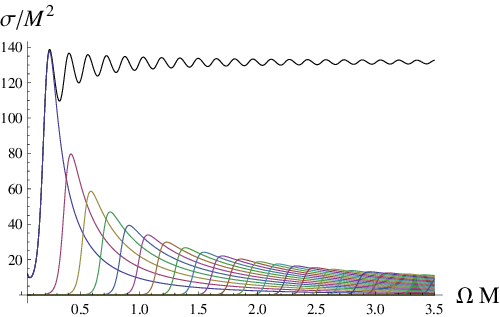}
\caption{Partial and total electromagnetic absorption cross sections for the DBI-supported regular black hole at $M=1$, obtained by summing the first 50 multipoles. Left panel: absorption cross sections for $a=0.2$. Right panel: absorption cross sections for $a=3$. In each panel, the partial absorption cross sections are shown together with the corresponding total absorption cross section.}
\label{fig:electromagnetic_absorption_compare}
\end{figure*}

\added{The absorption plots mirror the same barrier deformation seen in the grey-body factors. For all three spins, increasing the regularity parameter shifts the maxima of the dominant partial waves toward smaller frequencies, because the lower barriers allow the corresponding channels to transmit earlier. At the same time, the first few partial waves become more important, so the total absorption cross section is enhanced over essentially the whole frequency range displayed. Successive multipoles peak at progressively larger $\omega$ and with smaller height, reflecting the stronger centrifugal suppression of high-angular-momentum channels.}

\added{The scalar sector shows the strongest low-frequency enhancement because the $\ell=0$ mode contributes already at threshold. Consequently the total scalar absorption cross section remains large as $\omega\to0$ and is substantially higher for $a=3$ than for $a=0.2$, in agreement with the larger horizon area of the more regular solution. In the Dirac and electromagnetic sectors the low-frequency response is more strongly suppressed, so the first few partial waves appear as isolated resonant humps before the higher multipoles build up the total curve. In all three sectors the sums over partial waves produce damped oscillations around a nearly constant high-frequency plateau, which is the familiar approach to the geometric-optics capture cross section. The higher plateau obtained for larger $a$ shows that the DBI regularity scale enhances not only the onset of transmission but also the overall absorption efficiency at intermediate and high frequencies.}

\subsection{Geometric-optics estimate of the Hawking emission}

The total absorption cross section also determines the Hawking energy spectrum. For bosonic and fermionic fields one may write \cite{Hawking:1975vcx}
\begin{equation}\label{eq:hawking_flux}
\frac{d^2E}{dt\,d\omega}=\frac{\omega^3}{2\pi^2}\frac{\sigma_{\rm abs}(\omega)}{\exp(\omega/T_H)\mp 1},
\end{equation}
where the minus sign applies to bosons, the plus sign to fermions, and the Hawking temperature is 
\begin{equation}
T_H=\frac{f'(r_h)}{4\pi},
\end{equation}
with $r_h$ denoting the event-horizon radius. A full evaluation of Eq.~(\ref{eq:hawking_flux}) requires the exact frequency-dependent total absorption cross section in each spin sector. As a compact complement to the present scattering analysis, however, it is useful to consider the geometric-optics approximation
\begin{equation}
\sigma_{\rm abs}(\omega)\approx \sigma_{\rm geo}=\pi b_c^2,\qquad b_c^2=\frac{R^2(r_c)}{f(r_c)},
\end{equation}
where $r_c=3M$ is the null circular orbit. In this limit the integrated power becomes $P_{\rm geo}=\pi^2\sigma_{\rm geo}T_H^4/30$ for bosons and $P_{\rm geo}=7\pi^2\sigma_{\rm geo}T_H^4/240$ for fermions.

\begin{figure*}[p]
\centering
\includegraphics[width=\textwidth]{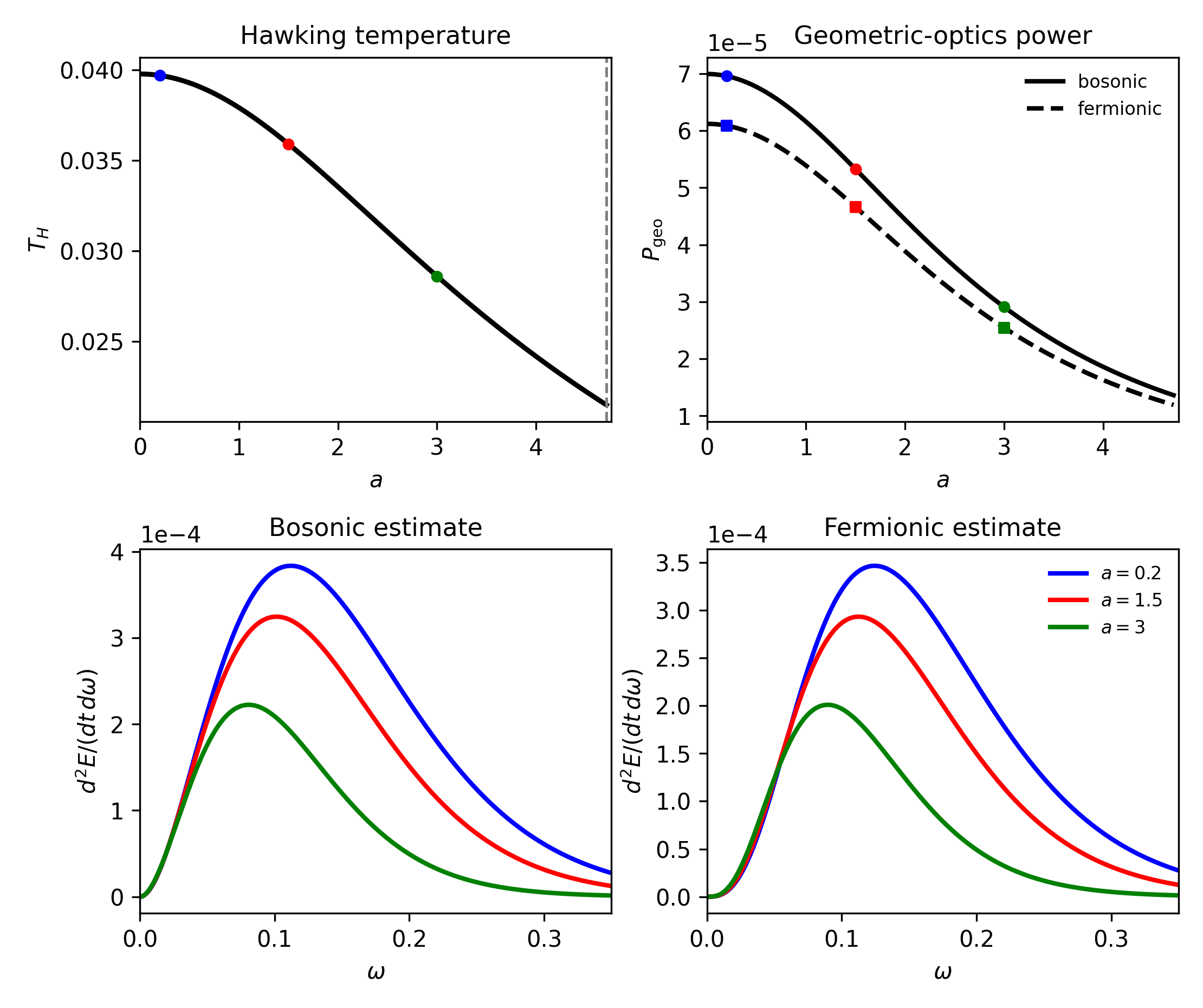}
\caption{Geometric-optics estimate of the Hawking emission in the DBI-supported regular black-hole branch. Top left: Hawking temperature $T_H$ as a function of the regularity parameter $a$. Top right: integrated geometric-optics power $P_{\rm geo}$ for bosonic (solid) and fermionic (dashed) statistics. Bottom panels: corresponding energy spectra $d^2E/(dt\,d\omega)$ for the representative values $a=0.2$ (blue), $a=1.5$ (red), and $a=3$ (green). The vertical dashed line marks the endpoint of the black-hole branch.}
\label{fig:hawking_geo}
\end{figure*}

Figure~\ref{fig:hawking_geo} summarizes the competition between cooling and the increase of the capture cross section. The Hawking temperature decreases monotonically as $a$ grows, whereas the geometric-optics capture cross section increases because the critical impact parameter becomes larger. For the representative values used throughout the grey-body-factor analysis, the enhancement of the capture cross section is not strong enough to compensate the thermal suppression proportional to $T_H^4$, so the total geometric-optics power decreases for both bosonic and fermionic statistics as the regularity parameter increases.

The spectral distributions show the complementary frequency-space picture. As $a$ increases, the maxima of the bosonic and fermionic spectra move to lower frequencies and their peak heights decrease. Thus the same regularity scale that enhances the low-frequency transmission probabilities also softens the Hawking output by lowering the temperature. A full partial-wave calculation based on the exact spin-dependent absorption cross sections would refine these estimates, especially in the low-frequency regime where the geometric-optics approximation is least accurate, but the present plot already captures the main competition between increased transparency and reduced temperature.

\clearpage

\section{Summary and outlook}\label{sec:summary}

In this paper we have studied grey-body factors and absorption cross sections for massless scalar, electromagnetic, and Dirac fields propagating in the exact asymptotically flat regular black-hole geometry supported by a phantom Dirac--Born--Infeld scalar. We have also supplemented this scattering analysis with a compact geometric-optics estimate of the corresponding Hawking emission. In the black-hole branch the effective potentials in all three sectors are single barriers that vanish at the horizon and at spatial infinity, which makes the scattering problem particularly suitable for a semi-analytic treatment. This allowed us to compute the transmission probabilities by the 6th-order WKB method and to compare them directly with the recent correspondence that reconstructs grey-body factors from the lowest quasinormal frequencies.

The physical picture that emerges from the results is clear and internally consistent. Increasing the regularity parameter $a$ lowers and broadens the effective barriers, so the onset of transmission is shifted to smaller frequencies and the grey-body factors increase for fixed spin and multipole number. In the bosonic sectors, raising the angular number from $\ell=1$ to $\ell=2$ produces the opposite effect, because the larger centrifugal contribution makes the barrier higher and delays the transition to efficient transmission. The Dirac mode with $\ell=3/2$ follows the same monotonic trend with $a$, confirming that the DBI regularity scale enhances transparency of the black hole in all three massless sectors.

The comparison between the direct WKB grey-body factors and the QNM/GBF correspondence shows that the latter is fulfilled with good quantitative accuracy already for the representative low multipoles considered here. The discrepancy is concentrated in the transition region where $0<\Gamma<1$, while it becomes negligible in the asymptotic low- and high-frequency regimes where the transmission coefficients saturate. For the displayed modes the maximal absolute difference remains at the level of $10^{-2}$ or smaller, and for the $\ell=2$ scalar and electromagnetic modes it is already reduced to a few $10^{-3}$. This supports the interpretation that, for single-barrier potentials, the same local geometry of the peak controls both the ringdown spectrum and the scattering characteristics even beyond the strict eikonal limit.

The absorption cross sections provide the complementary observable manifestation of the same barrier deformation. For all three spins, increasing $a$ shifts the dominant partial-wave peaks to lower frequencies and increases the total absorption cross section. The scalar field exhibits the strongest low-frequency absorption because the $\ell=0$ mode contributes already at threshold, whereas the Dirac and electromagnetic sectors remain more strongly suppressed near $\omega=0$. At higher frequencies the total cross sections oscillate around a nearly constant plateau associated with the geometric-optics capture regime, and this plateau is higher for larger values of $a$, again showing that the regularity scale enhances the overall absorptive efficiency of the DBI-supported black hole.

Combining the absorption analysis with a compact geometric-optics estimate of the Hawking emission reveals an instructive competition between transparency and temperature. Although larger $a$ enhances the capture cross section, this effect is more than compensated by the monotonic decrease of the Hawking temperature, so the total bosonic and fermionic power is reduced and the spectral peaks shift toward lower frequencies. In this sense the DBI regularity scale makes the black hole more transparent as a scatterer while rendering its thermal output softer and weaker in the geometric-optics approximation.

Several natural extensions follow from the present results. A first step would be to replace the present geometric-optics estimate by a full partial-wave Hawking-emission calculation based on the exact spin-dependent absorption cross sections and to implement systematically the improved two-mode version of the QNM/GBF correspondence in order to quantify the gain in accuracy for the lowest multipoles. It would also be interesting to extend the analysis to the extremal-remnant and horizonless branches of the same solution, and to include gravitational perturbations, for which the interplay between regularity, quasinormal ringing, scattering, and evaporation may reveal additional signatures of the underlying matter-supported geometry.

\begin{acknowledgments}
The author would like to acknowledge R. A. Konoplya for very helpful discussions.
\end{acknowledgments}

\bibliography{bibliography_gbf}

@article{Karmakar:2023cwg,
    author = "Karmakar, Ronit and Goswami, Umananda Dev",
    title = "{Quasinormal modes, temperatures and greybody factors of black holes in a generalized Rastall gravity theory}",
    eprint = "2310.18594",
    archivePrefix = "arXiv",
    primaryClass = "gr-qc",
    doi = "10.1088/1402-4896/ad350e",
    journal = "Phys. Scripta",
    volume = "99",
    number = "5",
    pages = "055003",
    year = "2024"
}

@article{Lutfuoglu:2026gey,
    author = {L{\"u}tf{\"u}o{\u{g}}lu, Bekir Can and Rayimbaev, Javlon and Murodov, Sardor and Kurbanov, Jakhongir and Matyoqubov, Muhammad},
    title = "{A First-Order Eikonal Framework for Quasinormal Modes, Shadows, Strong Lensing, and Grey-Body Factors in a Scalarized Black-Hole Metric}",
    eprint = "2604.14999",
    archivePrefix = "arXiv",
    primaryClass = "gr-qc",
    month = "4",
    year = "2026"
}

@article{Malik:2024sxv,
    author = "Malik, Zainab",
    title = "{Quasinormal Modes of Dilaton Black Holes: Analytic Approximations}",
    eprint = "2409.09872",
    archivePrefix = "arXiv",
    primaryClass = "gr-qc",
    doi = "10.1007/s10773-024-05660-5",
    journal = "Int. J. Theor. Phys.",
    volume = "63",
    number = "5",
    pages = "128",
    year = "2024"
}

@article{Skvortsova:2026jtx,
    author = "Skvortsova, Milena",
    title = "{Massive scalar quasinormal modes of an asymptotically flat regular black hole supported by a phantom Dirac--Born--Infeld field}",
    eprint = "2604.25471",
    archivePrefix = "arXiv",
    primaryClass = "gr-qc",
    month = "4",
    year = "2026"
}

@article{Lutfuoglu:2026boa,
    author = {L{\"u}tf{\"u}o{\u{g}}lu, Bekir Can},
    title = "{Scalar, electromagnetic, and Dirac perturbations of a regular black hole supported by primordial dark matter}",
    eprint = "2604.24349",
    archivePrefix = "arXiv",
    primaryClass = "gr-qc",
    month = "4",
    year = "2026"
}

@article{Saka:2025xxl,
    author = "Saka, Erdin{\c{c}} Ula{\c{s}}",
    title = "{Regular black hole sourced by the Dehnen-type distribution of matter: The sound of the event horizon}",
    eprint = "2512.08904",
    archivePrefix = "arXiv",
    primaryClass = "gr-qc",
    month = "12",
    year = "2025"
}

@article{Hawking:1975vcx,
    author = "Hawking, S. W.",
    editor = "Gibbons, G. W. and Hawking, S. W.",
    title = "{Particle Creation by Black Holes}",
    doi = "10.1007/BF02345020",
    journal = "Commun. Math. Phys.",
    volume = "43",
    pages = "199--220",
    year = "1975",
    note = "[Erratum: Commun.Math.Phys. 46, 206 (1976)]"
}

@article{Stuchlik:2025mjj,
    author = "Stuchl{\'\i}k, Z. and Zhidenko, A.",
    title = {{Non-oscillatory gravitational quasinormal modes of Reissner-Nordstr{\"o}m-de Sitter spacetime}},
    eprint = "2506.09829",
    archivePrefix = "arXiv",
    primaryClass = "gr-qc",
    month = "6",
    year = "2025"
}

@article{Malik:2024iky,
    author = "Malik, Zainab",
    title = "{Analytic Expressions for Quasinormal Modes in Einstein{\textendash}Aether Theory}",
    doi = "10.1134/S0202289325700598",
    journal = "Grav. Cosmol.",
    volume = "32",
    number = "1",
    pages = "122--134",
    year = "2026"
}

@article{Lutfuoglu:2025ohb,
    author = {L{\"u}tf{\"u}o{\u{g}}lu, B. C.},
    title = "{Quasinormal modes and gray-body factors for gravitational perturbations in asymptotically safe gravity}",
    eprint = "2505.06966",
    archivePrefix = "arXiv",
    primaryClass = "gr-qc",
    doi = "10.1140/epjc/s10052-026-15290-2",
    journal = "Eur. Phys. J. C",
    volume = "86",
    number = "1",
    pages = "39",
    year = "2026"
}

@article{Brill:1957fx,
  author  = {D. R. Brill and J. A. Wheeler},
  title   = {Interaction of Neutrinos and Gravitational Fields},
  journal = {Rev. Mod. Phys.},
  volume  = {29},
  pages   = {465--479},
  year    = {1957},
  doi     = {10.1103/RevModPhys.29.465}
}

@incollection{Ruffini:1973,
  author    = {R. Ruffini},
  title     = { },
  booktitle = {Black Holes: Les Astres Occlus},
  publisher = {Gordon and Breach},
  year      = {1973}
}

@article{Konoplya:2025ixm,
    author = "Konoplya, Roman A. and Pappas, Thomas D.",
    title = "{Dirty black holes, clean signals: near-horizon vs.~environmental effects on grey-body factors and Hawking radiation}",
    eprint = "2507.01954",
    archivePrefix = "arXiv",
    primaryClass = "gr-qc",
    doi = "10.1088/1475-7516/2026/02/038",
    journal = "JCAP",
    volume = "02",
    pages = "038",
    year = "2026"
}

@article{Page:1976df,
    author = "Page, Don N.",
    title = "{Particle Emission Rates from a Black Hole: Massless Particles from an Uncharged, Nonrotating Hole}",
    doi = "10.1103/PhysRevD.13.198",
    journal = "Phys. Rev. D",
    volume = "13",
    pages = "198--206",
    year = "1976"
}

@article{Page:1976ki,
    author = "Page, Don N.",
    title = "{Particle Emission Rates from a Black Hole. 2. Massless Particles from a Rotating Hole}",
    doi = "10.1103/PhysRevD.14.3260",
    journal = "Phys. Rev. D",
    volume = "14",
    pages = "3260--3273",
    year = "1976"
}

@article{Bonanno:2025dry,
    author = "Bonanno, Alfio M. and Konoplya, Roman A. and Oglialoro, Giovanni and Spina, Andrea",
    title = "{Regular black holes from proper-time flow in quantum gravity and their quasinormal modes, shadow and Hawking radiation}",
    eprint = "2509.12469",
    archivePrefix = "arXiv",
    primaryClass = "gr-qc",
    doi = "10.1088/1475-7516/2025/12/042",
    journal = "JCAP",
    volume = "12",
    pages = "042",
    year = "2025"
}

@article{Konoplya:2025uta,
    author = "Konoplya, Roman. A. and Zhidenko, Alexander",
    title = "{Convergence of Higher-Curvature Expansions Near the Horizon: Hawking Radiation from Regular Black Holes}",
    eprint = "2507.22660",
    archivePrefix = "arXiv",
    primaryClass = "gr-qc",
    doi = "10.53941/ijgtp.2025.100005",
    journal = "Int. J. Grav. Theor. Phys.",
    volume = "1",
    number = "1",
    pages = "5",
    year = "2025"
}

@article{Kanti:2002nr,
    author = "Kanti, Panagiota and March-Russell, John",
    title = "{Calculable corrections to brane black hole decay. 1. The scalar case}",
    eprint = "hep-ph/0203223",
    archivePrefix = "arXiv",
    reportNumber = "CERN-TH-2002-014",
    doi = "10.1103/PhysRevD.66.024023",
    journal = "Phys. Rev. D",
    volume = "66",
    pages = "024023",
    year = "2002"
}

@article{Konoplya:2019ppy,
    author = "Konoplya, R. A. and Zinhailo, A. F.",
    title = "{Hawking radiation of non-Schwarzschild black holes in higher derivative gravity: a crucial role of grey-body factors}",
    eprint = "1904.05341",
    archivePrefix = "arXiv",
    primaryClass = "gr-qc",
    doi = "10.1103/PhysRevD.99.104060",
    journal = "Phys. Rev. D",
    volume = "99",
    number = "10",
    pages = "104060",
    year = "2019"
}

@article{Konoplya:2024vuj,
    author = "Konoplya, R. A. and Zhidenko, A.",
    title = "{Correspondence between grey-body factors and quasinormal frequencies for rotating black holes}",
    eprint = "2408.11162",
    archivePrefix = "arXiv",
    primaryClass = "gr-qc",
    doi = "10.1016/j.physletb.2025.139288",
    journal = "Phys. Lett. B",
    volume = "861",
    pages = "139288",
    year = "2025"
}

@article{Berti:2009kk,
    author = "Berti, Emanuele and Cardoso, Vitor and Starinets, Andrei O.",
    title = "{Quasinormal modes of black holes and black branes}",
    eprint = "0905.2975",
    archivePrefix = "arXiv",
    primaryClass = "gr-qc",
    doi = "10.1088/0264-9381/26/16/163001",
    journal = "Class. Quant. Grav.",
    volume = "26",
    pages = "163001",
    year = "2009"
}

@article{Breton:2017hwe,
    author = "Breton, Nora and Clark, Tyler and Fernando, Sharmanthie",
    title = "{Quasinormal modes and absorption cross-sections of Born-Infeld-de Sitter black holes}",
    eprint = "1703.10070",
    archivePrefix = "arXiv",
    primaryClass = "gr-qc",
    doi = "10.1142/S0218271817501127",
    journal = "Int. J. Mod. Phys. D",
    volume = "26",
    number = "10",
    pages = "1750112",
    year = "2017"
}

@article{Cai:2021ele,
    author = "Cai, Xin-Chang and Miao, Yan-Gang",
    title = "{Quasinormal modes and shadows of a new family of Ayón-Beato-García black holes}",
    eprint = "2104.09725",
    archivePrefix = "arXiv",
    primaryClass = "gr-qc",
    doi = "10.1103/PhysRevD.103.124050",
    journal = "Phys. Rev. D",
    volume = "103",
    number = "12",
    pages = "124050",
    year = "2021"
}

@article{Dubinsky:2025wns,
    author = "Dubinsky, Alexey",
    title = "{Long-lived Modes and Grey-body Factors of Massive Fields in Quantum-corrected (Hayward) Black Holes}",
    eprint = "2511.00778",
    archivePrefix = "arXiv",
    primaryClass = "gr-qc",
    doi = "10.1007/s10773-026-06274-9",
    journal = "Int. J. Theor. Phys.",
    volume = "65",
    number = "2",
    pages = "45",
    year = "2026"
}

@article{Fernando:2012yw,
    author = "Fernando, Sharmanthie and Correa, Juan",
    title = "{Quasinormal Modes of Bardeen Black Hole: Scalar Perturbations}",
    eprint = "1208.5442",
    archivePrefix = "arXiv",
    primaryClass = "gr-qc",
    doi = "10.1103/PhysRevD.86.064039",
    journal = "Phys. Rev. D",
    volume = "86",
    pages = "064039",
    year = "2012"
}

@article{Flachi:2012nv,
    author = "Flachi, Antonino and Lemos, Jose P. S.",
    title = "{Quasinormal modes of regular black holes}",
    eprint = "1211.6212",
    archivePrefix = "arXiv",
    primaryClass = "gr-qc",
    doi = "10.1103/PhysRevD.87.024034",
    journal = "Phys. Rev. D",
    volume = "87",
    number = "2",
    pages = "024034",
    year = "2013"
}

@article{Gingrich:2024tuf,
    author = "Gingrich, Douglas M.",
    title = "{Quasinormal modes of a nonsingular spherically symmetric black hole effective model with holonomy corrections}",
    eprint = "2404.04447",
    archivePrefix = "arXiv",
    primaryClass = "gr-qc",
    month = "4",
    year = "2024"
}

@article{Guo:2024jhg,
    author = "Guo, Yang and Xie, Hao and Miao, Yan-Gang",
    title = "{Signal of phase transition hidden in quasinormal modes of regular AdS black holes, arXiv: 2402.10406}",
    eprint = "2402.10406",
    archivePrefix = "arXiv",
    primaryClass = "gr-qc",
    month = "2",
    year = "2024"
}

@article{Held:2019xde,
    author = "Held, Aaron and Gold, Roman and Eichhorn, Astrid",
    title = "{Asymptotic safety casts its shadow}",
    eprint = "1904.07133",
    archivePrefix = "arXiv",
    primaryClass = "gr-qc",
    doi = "10.1088/1475-7516/2019/06/029",
    journal = "JCAP",
    volume = "06",
    pages = "029",
    year = "2019"
}

@article{Huang:2023aet,
    author = "Huang, Bai-Hao and Hu, Han-Wen and Zhao, Liu",
    title = "{Thermodynamics for regular black holes as intermediate thermodynamic states and quasinormal frequencies}",
    eprint = "2311.12286",
    archivePrefix = "arXiv",
    primaryClass = "gr-qc",
    doi = "10.1088/1475-7516/2024/03/053",
    journal = "JCAP",
    volume = "03",
    pages = "053",
    year = "2024"
}

@article{Ishihara:2008re,
    author = "Ishihara, Hideki and Kimura, Masashi and Konoplya, Roman A. and Murata, Keiju and Soda, Jiro and Zhidenko, Alexander",
    title = "{Evolution of perturbations of squashed Kaluza-Klein black holes: escape from instability}",
    eprint = "0802.0655",
    archivePrefix = "arXiv",
    primaryClass = "hep-th",
    doi = "10.1103/PhysRevD.77.084019",
    journal = "Phys. Rev. D",
    volume = "77",
    pages = "084019",
    year = "2008"
}

@article{Iyer:1986np,
    author = "Iyer, Sai and Will, Clifford M.",
    title = "{Black Hole Normal Modes: A {WKB} Approach. 1. Foundations and Application of a Higher Order {WKB} Analysis of Potential Barrier Scattering}",
    reportNumber = "Print-86-1482 (WASH. U., ST. LOUIS)",
    doi = "10.1103/PhysRevD.35.3621",
    journal = "Phys. Rev. D",
    volume = "35",
    pages = "3621",
    year = "1987"
}

@article{Jawad:2020hju,
    author = "Jawad, Abdul and Yasir, Muhammad and Rani, Shamaila",
    title = "{Joule–Thomson expansion and quasinormal modes of regular non-minimal magnetic black hole}",
    doi = "10.1142/S0217732320502983",
    journal = "Mod. Phys. Lett. A",
    volume = "35",
    number = "36",
    pages = "2050298",
    year = "2020"
}

@article{Jusufi:2020odz,
    author = {Jusufi, Kimet and Azreg-Ainou, Mustapha and Jamil, Mubasher and Wei, Shao-Wen and Wu, Qiang and Wang, Anzhong},
    title = "{Quasinormal modes, quasiperiodic oscillations, and the shadow of rotating regular black holes in nonminimally coupled Einstein-Yang-Mills theory}",
    eprint = "2008.08450",
    archivePrefix = "arXiv",
    primaryClass = "gr-qc",
    doi = "10.1103/PhysRevD.103.024013",
    journal = "Phys. Rev. D",
    volume = "103",
    number = "2",
    pages = "024013",
    year = "2021"
}

@article{Kokkotas:1999bd,
    author = "Kokkotas, Kostas D. and Schmidt, Bernd G.",
    title = "{Quasinormal modes of stars and black holes}",
    eprint = "gr-qc/9909058",
    archivePrefix = "arXiv",
    doi = "10.12942/lrr-1999-2",
    journal = "Living Rev. Rel.",
    volume = "2",
    pages = "2",
    year = "1999"
}

@article{Konoplya:2003ii,
    author = "Konoplya, R. A.",
    title = "{Quasinormal behavior of the d-dimensional Schwarzschild black hole and higher order WKB approach}",
    eprint = "gr-qc/0303052",
    archivePrefix = "arXiv",
    doi = "10.1103/PhysRevD.68.024018",
    journal = "Phys. Rev. D",
    volume = "68",
    pages = "024018",
    year = "2003"
}

@article{Konoplya:2010vz,
    author = "Konoplya, R. A. and Zhidenko, A.",
    title = "{Long life of Gauss-Bonnet corrected black holes}",
    eprint = "1004.3772",
    archivePrefix = "arXiv",
    primaryClass = "hep-th",
    doi = "10.1103/PhysRevD.82.084003",
    journal = "Phys. Rev. D",
    volume = "82",
    pages = "084003",
    year = "2010"
}

@article{Konoplya:2011qq,
    author = "Konoplya, R. A. and Zhidenko, A.",
    title = "{Quasinormal modes of black holes: From astrophysics to string theory}",
    eprint = "1102.4014",
    archivePrefix = "arXiv",
    primaryClass = "gr-qc",
    doi = "10.1103/RevModPhys.83.793",
    journal = "Rev. Mod. Phys.",
    volume = "83",
    pages = "793--836",
    year = "2011"
}

@article{Konoplya:2017lhs,
    author = "Konoplya, R. A. and Zhidenko, A.",
    title = "{The portrait of eikonal instability in Lovelock theories}",
    eprint = "1705.01656",
    archivePrefix = "arXiv",
    primaryClass = "hep-th",
    doi = "10.1088/1475-7516/2017/05/050",
    journal = "JCAP",
    volume = "05",
    pages = "050",
    year = "2017"
}

@article{Konoplya:2017ymp,
    author = "Konoplya, R. A. and Zhidenko, A.",
    title = "{Eikonal instability of Gauss-Bonnet–(anti-)–de Sitter black holes}",
    eprint = "1701.01652",
    archivePrefix = "arXiv",
    primaryClass = "hep-th",
    doi = "10.1103/PhysRevD.95.104005",
    journal = "Phys. Rev. D",
    volume = "95",
    number = "10",
    pages = "104005",
    year = "2017"
}

@article{Konoplya:2017zwo,
    author = "Konoplya, R. A. and Zhidenko, A.",
    title = "{Quasinormal modes of Gauss-Bonnet-AdS black holes: towards holographic description of finite coupling}",
    eprint = "1705.07732",
    archivePrefix = "arXiv",
    primaryClass = "hep-th",
    doi = "10.1007/JHEP09(2017)139",
    journal = "JHEP",
    volume = "09",
    pages = "139",
    year = "2017"
}

@article{Konoplya:2020pbh,
    author = "Konoplya, R. A. and Zhidenko, A.",
    title = "{General parametrization of black holes: The only parameters that matter}",
    eprint = "2001.06100",
    archivePrefix = "arXiv",
    primaryClass = "gr-qc",
    doi = "10.1103/PhysRevD.101.124004",
    journal = "Phys. Rev. D",
    volume = "101",
    number = "12",
    pages = "124004",
    year = "2020"
}

@article{Konoplya:2022hll,
    author = "Konoplya, R. A. and Zinhailo, A. F. and Kunz, J. and Stuchlik, Z. and Zhidenko, A.",
    title = "{Quasinormal ringing of regular black holes in asymptotically safe gravity: the importance of overtones}",
    eprint = "2206.14714",
    archivePrefix = "arXiv",
    primaryClass = "gr-qc",
    doi = "10.1088/1475-7516/2022/10/091",
    journal = "JCAP",
    volume = "10",
    pages = "091",
    year = "2022"
}

@article{Konoplya:2023ppx,
    author = "Konoplya, R. A.",
    title = "{Quasinormal modes and grey-body factors of regular black holes with a scalar hair from the Effective Field Theory}",
    eprint = "2305.09187",
    archivePrefix = "arXiv",
    primaryClass = "gr-qc",
    doi = "10.1088/1475-7516/2023/07/001",
    journal = "JCAP",
    volume = "07",
    pages = "001",
    year = "2023"
}

@article{Li:2014fka,
    author = "Li, Jin and Lin, Kai and Yang, Nan",
    title = "{Nonlinear electromagnetic quasinormal modes and Hawking radiation of a regular black hole with magnetic charge}",
    eprint = "1409.5988",
    archivePrefix = "arXiv",
    primaryClass = "gr-qc",
    doi = "10.1140/epjc/s10052-015-3347-3",
    journal = "Eur. Phys. J. C",
    volume = "75",
    number = "3",
    pages = "131",
    year = "2015"
}

@article{Lin:2013ofa,
    author = "Lin, Kai and Li, Jin and Yang, Shuzheng",
    title = "{Quasinormal Modes of Hayward Regular Black Hole}",
    doi = "10.1007/s10773-013-1682-4",
    journal = "Int. J. Theor. Phys.",
    volume = "52",
    pages = "3771--3778",
    year = "2013"
}

@article{Lopez:2022uie,
    author = "Lopez, L. A. and Ramírez, Valeria",
    title = "{Quasi-normal modes of a Generic-class of magnetically charged regular black hole: scalar and electromagnetic perturbations}",
    eprint = "2205.10166",
    archivePrefix = "arXiv",
    primaryClass = "gr-qc",
    doi = "10.1140/epjp/s13360-023-03735-6",
    journal = "Eur. Phys. J. Plus",
    volume = "138",
    number = "2",
    pages = "120",
    year = "2023"
}

@article{Lutfuoglu:2025eik,
    author = {L{ü}tf{ü}o{\u{g}}lu, Bekir Can},
    title = "{Grey-Body Factors and Absorption Cross-Sections~of Scalar and Dirac Fields in the Vicinity of Dilaton-De Sitter Black Hole}",
    eprint = "2510.10579",
    archivePrefix = "arXiv",
    primaryClass = "gr-qc",
    doi = "10.1002/prop.70074",
    journal = "Fortsch. Phys.",
    volume = "74",
    number = "1",
    pages = "e70074",
    year = "2026"
}

@article{Lutfuoglu:2025kqp,
    author = {L{ü}tf{ü}o{\u{g}}lu, B. C.},
    title = "{Long-lived quasinormal modes, grey-body factors and absorption cross-section of the black hole immersed in the Hernquist galactic halo}",
    eprint = "2510.25969",
    archivePrefix = "arXiv",
    primaryClass = "gr-qc",
    doi = "10.1016/j.physletb.2025.140082",
    journal = "Phys. Lett. B",
    volume = "872",
    pages = "140082",
    year = "2026"
}

@article{Lutfuoglu:2026gis,
    author = {L{ü}tf{ü}o{\u{g}}lu, Bekir Can},
    title = "{Long-lived quasinormal modes, shadows and particle motion in four-dimensional quasi-topological gravity}",
    eprint = "2603.10844",
    archivePrefix = "arXiv",
    primaryClass = "gr-qc",
    month = "3",
    year = "2026"
}

@article{Lutfuoglu:2026uzy,
    author = {L{ü}tf{ü}o{\u{g}}lu, Bekir Can},
    title = "{Quasinormal Modes of a Massive Scalar Field in 4D Einstein--Gauss--Bonnet Black Hole Spacetimes}",
    eprint = "2603.24424",
    archivePrefix = "arXiv",
    primaryClass = "gr-qc",
    month = "3",
    year = "2026"
}

@article{MahdavianYekta:2019pol,
    author = "Mahdavian Yekta, Davood and Karimabadi, Majid and Alavi, S. A.",
    title = "{Quasinormal modes for non-minimally coupled scalar fields in regular black hole spacetimes: Grey-body factors, area spectrum and shadow radius}",
    eprint = "1912.12017",
    archivePrefix = "arXiv",
    primaryClass = "hep-th",
    doi = "10.1016/j.aop.2021.168603",
    journal = "Annals Phys.",
    volume = "434",
    pages = "168603",
    year = "2021"
}

@article{Matyjasek:2017psv,
    author = "Matyjasek, Jerzy and Opala, Michał",
    title = "{Quasinormal modes of black holes. The improved semianalytic approach}",
    eprint = "1704.00361",
    archivePrefix = "arXiv",
    primaryClass = "gr-qc",
    doi = "10.1103/PhysRevD.96.024011",
    journal = "Phys. Rev. D",
    volume = "96",
    number = "2",
    pages = "024011",
    year = "2017"
}

@article{Pedraza:2021hzw,
    author = "Pedraza, Omar and López, L. A. and Arceo, R. and Cabrera-Munguia, I.",
    title = "{Quasinormal modes of the Hayward black hole surrounded by quintessence: Scalar, electromagnetic and gravitational perturbations}",
    eprint = "2111.06488",
    archivePrefix = "arXiv",
    primaryClass = "gr-qc",
    doi = "10.1142/S0217732322500572",
    journal = "Mod. Phys. Lett. A",
    volume = "37",
    number = "09",
    pages = "2250057",
    year = "2022"
}

@article{Parvez:2025dbi,
    author = "Parvez, Tausif and Shankaranarayanan, S.",
    title = "{Exact, non-singular black holes from a phantom DBI field as primordial dark matter}",
    eprint = "2511.14047",
    archivePrefix = "arXiv",
    primaryClass = "gr-qc",
    journal = "arXiv e-prints, accepted for publication in Phys. Rev. D.",
    year = "2025"
}

@article{Skvortsova:2024eqi,
    author = "Skvortsova, Milena",
    title = "{Long-lived quasinormal modes of regular and extreme black holes}",
    eprint = "2503.03650",
    archivePrefix = "arXiv",
    primaryClass = "gr-qc",
    doi = "10.1209/0295-5075/adaee2",
    journal = "EPL",
    volume = "149",
    number = "5",
    pages = "59001",
    year = "2025"
}

@article{Bolokhov:2025fto,
    author = "Bolokhov, S. V.",
    title = "{Quasinormal ringing of a regular black hole sourced by the Dehnen-type distribution of matter}",
    eprint = "2511.12859",
    archivePrefix = "arXiv",
    primaryClass = "gr-qc",
    doi = "10.1016/j.aop.2026.170416",
    journal = "Annals Phys.",
    volume = "488",
    pages = "170416",
    year = "2026"
}

@article{Wongjun:2019ydo,
    author = "Wongjun, Pitayuth and Chen, Chun-Hung and Nakarachinda, Ratchaphat",
    title = "{Quasinormal modes of a massless Dirac field in de Rham-Gabadadze-Tolley massive gravity}",
    eprint = "1910.05908",
    archivePrefix = "arXiv",
    primaryClass = "gr-qc",
    doi = "10.1103/PhysRevD.101.124033",
    journal = "Phys. Rev. D",
    volume = "101",
    number = "12",
    pages = "124033",
    year = "2020"
}

@article{Yang:2021cvh,
    author = "Yang, Yi and Liu, Dong and Xu, Zhaoyi and Xing, Yujia and Wu, Shurui and Long, Zheng-Wen",
    title = "{Echoes of novel black-bounce spacetimes}",
    eprint = "2107.06554",
    archivePrefix = "arXiv",
    primaryClass = "gr-qc",
    doi = "10.1103/PhysRevD.104.104021",
    journal = "Phys. Rev. D",
    volume = "104",
    number = "10",
    pages = "104021",
    year = "2021"
}

@article{Konoplya:2024gbf,
    author = "Konoplya, R. A. and Zhidenko, A.",
    title = "{Correspondence between grey-body factors and quasinormal modes}",
    eprint = "2406.11694",
    archivePrefix = "arXiv",
    primaryClass = "gr-qc",
    doi = "10.1088/1475-7516/2024/09/068",
    journal = "JCAP",
    volume = "09",
    pages = "068",
    year = "2024"
}

@article{Bolokhov:2026eqf,
    author = "Bolokhov, S. V.",
    title = "{Quasinormal Modes and Grey-Body Factors of Scalar, Electromagnetic and Dirac Fields Around Einasto-Supported Regular Black Holes}",
    eprint = "2603.22310",
    archivePrefix = "arXiv",
    primaryClass = "gr-qc",
    month = "3",
    year = "2026"
}

@article{Arbelaez:2026eaz,
    author = "Arbelaez, Juan Pablo",
    title = "{Grey-body factors of higher dimensional regular black holes in quasi-topological theories}",
    eprint = "2601.22340",
    archivePrefix = "arXiv",
    primaryClass = "gr-qc",
    month = "1",
    year = "2026"
}

@article{Arbelaez:2025gwj,
    author = "Arbelaez, Juan Pablo",
    title = "{Quasinormal spectra of higher dimensional regular black holes in theories with infinite curvature corrections}",
    eprint = "2509.25141",
    archivePrefix = "arXiv",
    primaryClass = "gr-qc",
    month = "9",
    year = "2025"
}

@article{Konoplya:2019hlu,
    author = "Konoplya, R. A. and Zhidenko, A. and Zinhailo, A. F.",
    title = "{Higher order WKB formula for quasinormal modes and grey-body factors: recipes for quick and accurate calculations}",
    eprint = "1904.10333",
    archivePrefix = "arXiv",
    primaryClass = "gr-qc",
    doi = "10.1088/1361-6382/ab2e25",
    journal = "Class. Quant. Grav.",
    volume = "36",
    pages = "155002",
    year = "2019"
}

@article{Abdalla:2005hu,
    author = "Abdalla, E. and Konoplya, R. A. and Molina, C.",
    title = "{Scalar field evolution in Gauss-Bonnet black holes}",
    eprint = "hep-th/0507100",
    archivePrefix = "arXiv",
    doi = "10.1103/PhysRevD.72.084006",
    journal = "Phys. Rev. D",
    volume = "72",
    pages = "084006",
    year = "2005"
}

@article{Fernando:2016ftj,
    author = "Fernando, Sharmanthie",
    title = "{Quasinormal modes of dilaton-de Sitter black holes: scalar perturbations}",
    eprint = "1601.06407",
    archivePrefix = "arXiv",
    primaryClass = "gr-qc",
    doi = "10.1007/s10714-016-2020-y",
    journal = "Gen. Rel. Grav.",
    volume = "48",
    number = "3",
    pages = "24",
    year = "2016"
}

@article{Bolokhov:2025egl,
    author = "Bolokhov, S. V. and Skvortsova, Milena",
    title = "{Gravitational quasinormal modes of the Hayward spacetime}",
    eprint = "2508.19989",
    archivePrefix = "arXiv",
    primaryClass = "gr-qc",
    doi = "10.1140/epjc/s10052-026-15624-0",
    journal = "Eur. Phys. J. C",
    volume = "86",
    number = "4",
    pages = "374",
    year = "2026"
}

@article{Konoplya:2005sy,
    author = "Konoplya, R. A. and Abdalla, Elcio",
    title = "{Scalar field perturbations of the Schwarzschild black hole in the Godel universe}",
    eprint = "hep-th/0503029",
    archivePrefix = "arXiv",
    doi = "10.1103/PhysRevD.71.084015",
    journal = "Phys. Rev. D",
    volume = "71",
    pages = "084015",
    year = "2005"
}

@article{Konoplya:2001ji,
    author = "Konoplya, R. A.",
    title = "{Quasinormal modes of the electrically charged dilaton black hole}",
    eprint = "gr-qc/0109096",
    archivePrefix = "arXiv",
    doi = "10.1023/A:1015347628961",
    journal = "Gen. Rel. Grav.",
    volume = "34",
    pages = "329--335",
    year = "2002"
}

@article{Konoplya:2006ar,
    author = "Konoplya, R. A. and Zhidenko, A.",
    title = "{Gravitational spectrum of black holes in the Einstein-Aether theory}",
    eprint = "hep-th/0611226",
    archivePrefix = "arXiv",
    doi = "10.1016/j.physletb.2007.03.018",
    journal = "Phys. Lett. B",
    volume = "648",
    pages = "236--239",
    year = "2007"
}

@article{Kokkotas:2010zd,
    author = "Kokkotas, K. D. and Konoplya, R. A. and Zhidenko, A.",
    title = "{Quasinormal modes, scattering and Hawking radiation of Kerr-Newman black holes in a magnetic field}",
    eprint = "1011.1843",
    archivePrefix = "arXiv",
    primaryClass = "gr-qc",
    doi = "10.1103/PhysRevD.83.024031",
    journal = "Phys. Rev. D",
    volume = "83",
    pages = "024031",
    year = "2011"
}

@article{Malik:2025erb,
    author = "Malik, Zainab",
    title = "{Grey-Body Factors for Scalar and Dirac Fields in the Euler-Heisenberg Electrodynamics}",
    eprint = "2509.15995",
    archivePrefix = "arXiv",
    primaryClass = "gr-qc",
    doi = "10.53941/ijgtp.2025.100006",
    journal = "Int. J. Grav. Theor. Phys.",
    volume = "1",
    number = "1",
    pages = "6",
    year = "2025"
}

@article{Pathrikar:2025gzu,
    author = "Pathrikar, Akshat",
    title = "{Quasinormal Ringing and Unruh-Verlinde Temperature of the Frolov Black Hole}",
    eprint = "2510.01376",
    archivePrefix = "arXiv",
    primaryClass = "gr-qc",
    doi = "10.53941/ijgtp.2026.100001",
    journal = "Int. J. Grav. Theor. Phys.",
    volume = "1",
    number = "1",
    pages = "1",
    year = "2026"
}

@article{Momennia:2018hsm,
    author = "Momennia, Mehrab and Hossein Hendi, Seyed and Soltani Bidgoli, Fatemeh",
    title = "{Stability and quasinormal modes of black holes in conformal Weyl gravity}",
    eprint = "1807.01792",
    archivePrefix = "arXiv",
    primaryClass = "hep-th",
    doi = "10.1016/j.physletb.2020.136028",
    journal = "Phys. Lett. B",
    volume = "813",
    pages = "136028",
    year = "2021"
}

@article{Guo:2020caw,
    author = "Guo, Yang and Miao, Yan-Gang",
    title = "{Scalar quasinormal modes of black holes in Einstein-Yang-Mills gravity}",
    eprint = "2005.07524",
    archivePrefix = "arXiv",
    primaryClass = "hep-th",
    doi = "10.1103/PhysRevD.102.064049",
    journal = "Phys. Rev. D",
    volume = "102",
    number = "6",
    pages = "064049",
    year = "2020"
}

@article{Konoplya:2009hv,
    author = "Konoplya, R. A. and Zhidenko, A.",
    title = "{Holographic conductivity of zero temperature superconductors}",
    eprint = "0909.2138",
    archivePrefix = "arXiv",
    primaryClass = "hep-th",
    doi = "10.1016/j.physletb.2010.02.048",
    journal = "Phys. Lett. B",
    volume = "686",
    pages = "199--206",
    year = "2010"
}

@article{Eniceicu:2019npi,
    author = "Eniceicu, Dan Stefan and Reece, Matthew",
    title = {{Quasinormal modes of charged fields in Reissner-Nordstr{\"o}m backgrounds by Borel-Pad{\'e} summation of Bender-Wu series}},
    eprint = "1912.05553",
    archivePrefix = "arXiv",
    primaryClass = "gr-qc",
    doi = "10.1103/PhysRevD.102.044015",
    journal = "Phys. Rev. D",
    volume = "102",
    number = "4",
    pages = "044015",
    year = "2020"
}

@article{Bolokhov:2025aqy,
    author = "Bolokhov, Sergei V. and Bronnikov, Kirill A. and Skvortsova, Milena V.",
    title = "{Stability Ranges of Magnetic Black Holes and Mirror (Topological) Stars in 5D Gravity}",
    eprint = "2507.14603",
    archivePrefix = "arXiv",
    primaryClass = "gr-qc",
    doi = "10.53941/ijgtp.2026.100002",
    journal = "Int. J. Grav. Theor. Phys.",
    volume = "2",
    number = "1",
    pages = "2",
    year = "2026"
}

@article{Bolokhov:2025lnt,
    author = "Bolokhov, S. V. and Skvortsova, Milena",
    title = "{Gravitational Quasinormal Modes and Grey-Body Factors of Bonanno{\textendash}Reuter Regular Black Holes}",
    eprint = "2507.07196",
    archivePrefix = "arXiv",
    primaryClass = "gr-qc",
    doi = "10.53941/ijgtp.2025.100003",
    journal = "Int. J. Grav. Theor. Phys.",
    volume = "1",
    number = "1",
    pages = "3",
    year = "2025"
}

@article{Skvortsova:2024msa,
    author = "Skvortsova, Milena",
    title = "{Quantum corrected black holes: testing the correspondence between grey-body factors and quasinormal modes}",
    eprint = "2411.06007",
    archivePrefix = "arXiv",
    primaryClass = "gr-qc",
    doi = "10.1140/epjc/s10052-025-14589-w",
    journal = "Eur. Phys. J. C",
    volume = "85",
    number = "8",
    pages = "854",
    year = "2025"
}

@article{Dubinsky:2026gcj,
    author = "Dubinsky, Alexey",
    title = "{From Ringdown to Lensing: Analytic Eikonal Modes of Quasi-Topological Regular Black Holes}",
    eprint = "2604.13613",
    archivePrefix = "arXiv",
    primaryClass = "gr-qc",
    month = "4",
    year = "2026"
}

@article{Malik:2024itg,
    author = "Malik, Zainab",
    title = "{Analytic quasinormal frequencies of the Casadio{\textendash}Fabbri{\textendash}Mazzacurati black hole}",
    doi = "10.1139/cjp-2024-0247",
    journal = "Can. J. Phys.",
    volume = "103",
    number = "9",
    pages = "800--811",
    year = "2025"
}

@article{Malik:2024tuf,
    author = "Malik, Zainab",
    title = "{Analytical QNMs of fields of various spin in the Hayward spacetime}",
    eprint = "2410.04306",
    archivePrefix = "arXiv",
    primaryClass = "gr-qc",
    reportNumber = "Research Gate Preprint: DOI:10.13140/RG.2.2.32496.06402",
    doi = "10.1209/0295-5075/ad7885",
    journal = "EPL",
    volume = "147",
    number = "6",
    pages = "69001",
    year = "2024"
}

@article{Malik:2024cgb,
    author = "Malik, Zainab",
    title = "{Correspondence between quasinormal modes and grey-body factors for massive fields in Schwarzschild-de~Sitter spacetime}",
    eprint = "2412.19443",
    archivePrefix = "arXiv",
    primaryClass = "gr-qc",
    doi = "10.1088/1475-7516/2025/04/042",
    journal = "JCAP",
    volume = "04",
    pages = "042",
    year = "2025"
}

@article{Malik:2025dxn,
    author = "Malik, Zainab",
    title = "{Gravitational Perturbations of the Hayward Spacetime and Testing the Correspondence between Quasinormal Modes and Grey-body Factors}",
    eprint = "2508.19178",
    archivePrefix = "arXiv",
    primaryClass = "gr-qc",
    doi = "10.1007/s10773-025-06198-w",
    journal = "Int. J. Theor. Phys.",
    volume = "64",
    number = "11",
    pages = "314",
    year = "2025"
}

@article{Malik:2026lfj,
    author = "Malik, Zainab",
    title = "{Analytic Expressions for Quasinormal Modes of a Regular Black Hole Sourced by a Dehnen-Type Halo}",
    eprint = "2603.18887",
    archivePrefix = "arXiv",
    primaryClass = "gr-qc",
    doi = "10.53941/ijgtp.2026.100003",
    journal = "Int. J. Grav. Theor. Phys.",
    volume = "2",
    number = "1",
    pages = "3",
    year = "2026"
}

@article{Carter1968HJ,
  author  = {Carter, Brandon},
  title   = {Hamilton-Jacobi and Schr{\"o}dinger Separable Solutions of Einstein's Equations},
  journal = {Communications in Mathematical Physics},
  volume  = {10},
  number  = {4},
  pages   = {280--310},
  year    = {1968},
  doi     = {10.1007/BF03399503}
}

@article{Carter1968Kerr,
  author  = {Carter, Brandon},
  title   = {Global Structure of the Kerr Family of Gravitational Fields},
  journal = {Physical Review},
  volume  = {174},
  number  = {5},
  pages   = {1559--1571},
  year    = {1968},
  doi     = {10.1103/PhysRev.174.1559}
}

@article{Konoplya:2018arm,
    author = "Konoplya, R. A. and Stuchl{\'\i}k, Z. and Zhidenko, A.",
    title = "{Axisymmetric black holes allowing for separation of variables in the Klein-Gordon and Hamilton-Jacobi equations}",
    eprint = "1801.07195",
    archivePrefix = "arXiv",
    primaryClass = "gr-qc",
    doi = "10.1103/PhysRevD.97.084044",
    journal = "Phys. Rev. D",
    volume = "97",
    number = "8",
    pages = "084044",
    year = "2018"
}

@article{Takahashi:2010gz,
    author = "Takahashi, Tomohiro and Soda, Jiro",
    title = "{Catastrophic Instability of Small Lovelock Black Holes}",
    eprint = "1008.1618",
    archivePrefix = "arXiv",
    primaryClass = "gr-qc",
    reportNumber = "KUNS-2288",
    doi = "10.1143/PTP.124.711",
    journal = "Prog. Theor. Phys.",
    volume = "124",
    pages = "711--729",
    year = "2010"
}

@article{Dotti:2004sh,
    author = "Dotti, Gustavo and Gleiser, Reinaldo J.",
    title = "{Gravitational instability of Einstein-Gauss-Bonnet black holes under tensor mode perturbations}",
    eprint = "gr-qc/0409005",
    archivePrefix = "arXiv",
    doi = "10.1088/0264-9381/22/1/L01",
    journal = "Class. Quant. Grav.",
    volume = "22",
    pages = "L1",
    year = "2005"
}

@article{Konoplya:2025hgp,
    author = "Konoplya, R. A. and Stashko, O. S.",
    title = "{Transition from regular black holes to wormholes in covariant effective quantum gravity: Scattering, quasinormal modes, and Hawking radiation}",
    eprint = "2502.05689",
    archivePrefix = "arXiv",
    primaryClass = "gr-qc",
    doi = "10.1103/PhysRevD.111.084031",
    journal = "Phys. Rev. D",
    volume = "111",
    number = "8",
    pages = "084031",
    year = "2025"
}

@article{Lutfuoglu:2025mqa,
    author = {L{\"u}tf{\"u}o{\u{g}}lu, Bekir Can and Shermatov, Abubakir and Rayimbaev, Javlon and Matyoqubov, Muhammad and Sirajiddin, Otaboyev},
    title = "{Gravitational spectra and wave propagation in regular black holes supported by a Dehnen Halo}",
    eprint = "2511.22366",
    archivePrefix = "arXiv",
    primaryClass = "gr-qc",
    doi = "10.1140/epjc/s10052-025-15234-2",
    journal = "Eur. Phys. J. C",
    volume = "85",
    number = "12",
    pages = "1484",
    year = "2025"
}

@article{Skvortsova:2024wly,
    author = "Skvortsova, Milena",
    title = "{Ringing of Extreme Regular Black Holes}",
    eprint = "2405.15807",
    archivePrefix = "arXiv",
    primaryClass = "gr-qc",
    doi = "10.1134/S020228932470018X",
    journal = "Grav. Cosmol.",
    volume = "30",
    number = "3",
    pages = "279--288",
    year = "2024"
}

@article{Bolokhov:2024otn,
    author = "Bolokhov, S. V. and Skvortsova, Milena",
    title = "{Correspondence between quasinormal modes and grey-body factors of spherically symmetric traversable wormholes}",
    eprint = "2412.11166",
    archivePrefix = "arXiv",
    primaryClass = "gr-qc",
    doi = "10.1088/1475-7516/2025/04/025",
    journal = "JCAP",
    volume = "04",
    pages = "025",
    year = "2025"
}

@article{Konoplya:2023aph,
    author = "Konoplya, R. A. and Stuchlik, Z. and Zhidenko, A. and Zinhailo, A. F.",
    title = "{Quasinormal modes of renormalization group improved Dymnikova regular black holes}",
    eprint = "2303.01987",
    archivePrefix = "arXiv",
    primaryClass = "gr-qc",
    doi = "10.1103/PhysRevD.107.104050",
    journal = "Phys. Rev. D",
    volume = "107",
    number = "10",
    pages = "104050",
    year = "2023"
}

@article{Konoplya:2025ect,
    author = "Konoplya, R. A. and Zhidenko, A.",
    title = "{Dark matter halo as a source of regular black-hole geometries}",
    eprint = "2511.03066",
    archivePrefix = "arXiv",
    primaryClass = "gr-qc",
    doi = "10.1103/7ptp-9j1t",
    journal = "Phys. Rev. D",
    volume = "113",
    number = "4",
    pages = "043011",
    year = "2026"
}

@article{Dubinsky:2024vbn,
    author = "Dubinsky, Alexey",
    title = "{Gray-body factors for gravitational and electromagnetic perturbations around Gibbons{\textendash}Maeda{\textendash}Garfinkle{\textendash}Horowitz{\textendash}Strominger black holes}",
    eprint = "2412.00625",
    archivePrefix = "arXiv",
    primaryClass = "gr-qc",
    doi = "10.1142/S0217732325501111",
    journal = "Mod. Phys. Lett. A",
    volume = "40",
    number = "28",
    pages = "2550111",
    year = "2025"
}

@article{Malik:2025qnr,
    author = "Malik, Zainab",
    title = "{Bonanno-Reuter regular black hole: quasi-resonances, grey-body factors and absorption cross-sections of a massive scalar field}",
    eprint = "2510.06689",
    archivePrefix = "arXiv",
    primaryClass = "gr-qc",
    month = "10",
    year = "2025"
}

@article{Malik:2024wvs,
    author = "Malik, Zainab",
    title = "{Quasinormal modes and grey-body factors of Morris-Thorne wormholes}",
    eprint = "2412.13385",
    archivePrefix = "arXiv",
    primaryClass = "gr-qc",
    month = "12",
    year = "2024"
}

@article{Konoplya:2026gim,
    author = "Konoplya, R. A.",
    title = "{Quasinormal modes of four-dimensional regular black holes in quasi-topological gravity: Overtones{\textquoteright} outburst via WKB method}",
    eprint = "2603.03189",
    archivePrefix = "arXiv",
    primaryClass = "gr-qc",
    doi = "10.1016/j.physletb.2026.140386",
    journal = "Phys. Lett. B",
    volume = "876",
    pages = "140386",
    year = "2026"
}

\end{document}